\documentclass{PoS}
\pdfoutput=1
\usepackage{rotating,booktabs,amsmath,amssymb, multicol,url}
\usepackage[numbers,sort&compress]{natbib}
\usepackage{natbibspacing}
\usepackage{ulem}
\usepackage{multicol}
\newcommand{\gGF}{\ensuremath{g_{\rm GF}^2} }

\newcommand{\gtGF}{\ensuremath{\widetilde g_{\rm GF}^2} }
\newcommand{\gtopt}{\ensuremath{\widetilde g_{\rm opt}^2} }

\title{Targeting the conformal window with 4+8 flavors}

\ShortTitle{Targeting the conformal window with 4+8 flavors}

\author{Rich Brower\\
        Department of Physics, Boston University, Boston, MA, USA\\   
        E-mail: \email{brower@bu.edu}}
\author{Anna Hasenfratz\\
        Department of Physics, University of Colorado Boulder, Boulder, CO, USA\\
        E-mail: \email{Anna.Hasenfratz@colorado.edu}}
\author{Claudio Rebbi\\
        Department of Physics, Boston University, Boston, MA, USA\\
        E-mail: \email{rebbi@bu.edu}}
\author{\speaker{Evan Weinberg}\\
        Department of Physics, Boston University, Boston, MA, USA\\
        E-mail: \email{weinbe2@bu.edu}}
\author{Oliver Witzel\footnote{Poster, Present address: School of Physics \& Astronomy, The University of Edinburgh, EH9 3FD, UK}\\
        Center for Computational Science, Boston University, Boston, MA, USA\\
        E-mail: \email{owitzel@bu.edu}}

\abstract{
We study the transition between spontaneous chiral symmetry breaking 
and conformal behavior in the $SU(3)$ theory with multiple fermion
flavors. Instead of the traditional approach of changing the number of flavors, we keep the number of fermions fixed but lift the
    mass of a subset, keeping the remaining fermions near to the massless chiral limit. This way we can interpolate continuously between the conformal 
   and chirally broken dynamics. In particular, we consider four light and eight heavy flavors and investigate the running/walking gauge coupling 
   and the low energy meson spectrum, including the $0^{++}$ iso-singlet scalar state in this system.  
Our preliminary data reveal an iso-singlet scalar that is considerably lighter than the pion at large fermion mass but becomes heavier at smaller  masses. This behavior is of particular phenomenological interest.
}

\FullConference{The 32nd International Symposium on Lattice Field Theory,\\
		23-28 June, 2014\\
		Columbia University New York, NY}

\begin{document}

\section{Introduction}
\label{Intro}

The experimental discovery of the Higgs boson at the LHC in 2012 \cite{Aad:2012tfa,Chatrchyan:2012ufa} contributed the missing piece of the electroweak sector of the Standard Model but so far we do not have experimental insight into the nature of electroweak symmetry breaking or the origin of the Higgs boson.   Given the current experimental constraints, there are two theoretically viable scenarios: supersymmetric extensions of the Standard Model and composite Higgs models. 
Both would solve the well-known problem that theories with self-interacting scalars require an ultraviolet completion, which is not contained in the Standard Model.
In this work we consider composite Higgs models that are based on a new, strongly coupled but chirally broken gauge-fermion sector where the Higgs boson is a fermionic $0^{++}$ bound state.  Composite Higgs models originate from ``Technicolor'' models~\cite{Weinberg:1979bn,Susskind:1978ms} which subsequently were supplemented by a mechanism to generate fermion masses (``Extended Technicolor'') \cite{Dimopoulos:1979es,Eichten:1979ah}  and later refined to ``Walking Technicolor'' to satisfy electroweak phenomenological constraints \cite{Yamawaki:1985zg,Appelquist:1991nm}. 

Any new gauge-fermion system predicts a plethora of new composite states which could be in tension with current experimental findings.  Hence any composite Higgs model faces the challenge to predict a bound scalar with the mass and properties of the Higgs boson while all other states are at sufficiently larger mass. A model exhibiting a walking behavior induced by a weakly broken  conformal symmetry is conjectured to be a promising candidate with the desired spectral properties. Such a system could be realized by a theory close to an infra-red fixed point (IRFP). In this respect, recent lattice results using the $SU(3)$ gauge group have triggered special interest: Independent groups have reported indications for an infrared fixed point in the theory with 12 fundamental flavors  \cite{Appelquist:2011dp,Hasenfratz:2011xn,Aoki:2012eq,DeGrand:2011cu,Cheng:2013xha,Cheng:2014jba,Lombardo:2014pda,Aoki:2013zsa}, and a low mass scalar has been observed in investigations with 8 fundamental flavors \cite{Aoki:2014oha}  or 2 sextet flavors \cite{Fodor:2014pqa}.

As exciting these results may be, the above considerations are merely a conjecture offering a possible explanation of electroweak symmetry breaking based on a strongly interacting gauge theory. Much more work is needed to (in)validate such a conjecture. Many models of colors, flavors, and fermion representation will need to be investigated because contrary to QCD, little is known experimentally. Like in the case of QCD, lattice simulations offer the only ab initio non-perturbative tool at present to study these non-Abelian gauge theories from first principles. 

Given a gauge group and fermion representation most studies choose the number of fermions  such that the system  is as close to the conformal window as possible.  Such models have the highest likelihood of exhibiting the phenomenologically desired properties i.e.~a walking scenario with a light mass iso-singlet scalar. The inherent difficulty is that the number of flavors is an integer and not a continuous variable. There is no guarantee that an integer value is close enough to the IRFP.  In this work we report on  a novel approach that avoids this inherent difficulty.

In our approach we split the masses of the fermions, keeping some of them near the chiral limit while lifting the  mass of the others~\cite{Brower:2014dfa}. Based on the evidence that an $SU(3)$ theory with 12 fundamental flavors exhibits an IRFP and the fact that an $SU(3)$ four flavor theory is chirally broken, QCD like, we chose to study a system with $N_\ell=4$ light and $N_h=8$ heavy flavors. We keep the $N_\ell$ light flavors near the chiral limit, i.e., $m_\ell \approx 0$ and give the $N_h$ heavy flavors a variable mass with $m_h\ge m_\ell$. This system interpolates between the  $N_h+N_\ell=12$ flavor  mass deformed conformal and the $N_\ell=4$ chirally broken models. Effectively we replace the discrete flavor number by a continuous mass parameter $m_h$. 

\begin{figure}[tb]
\begin{center}
    \includegraphics[width=8cm]{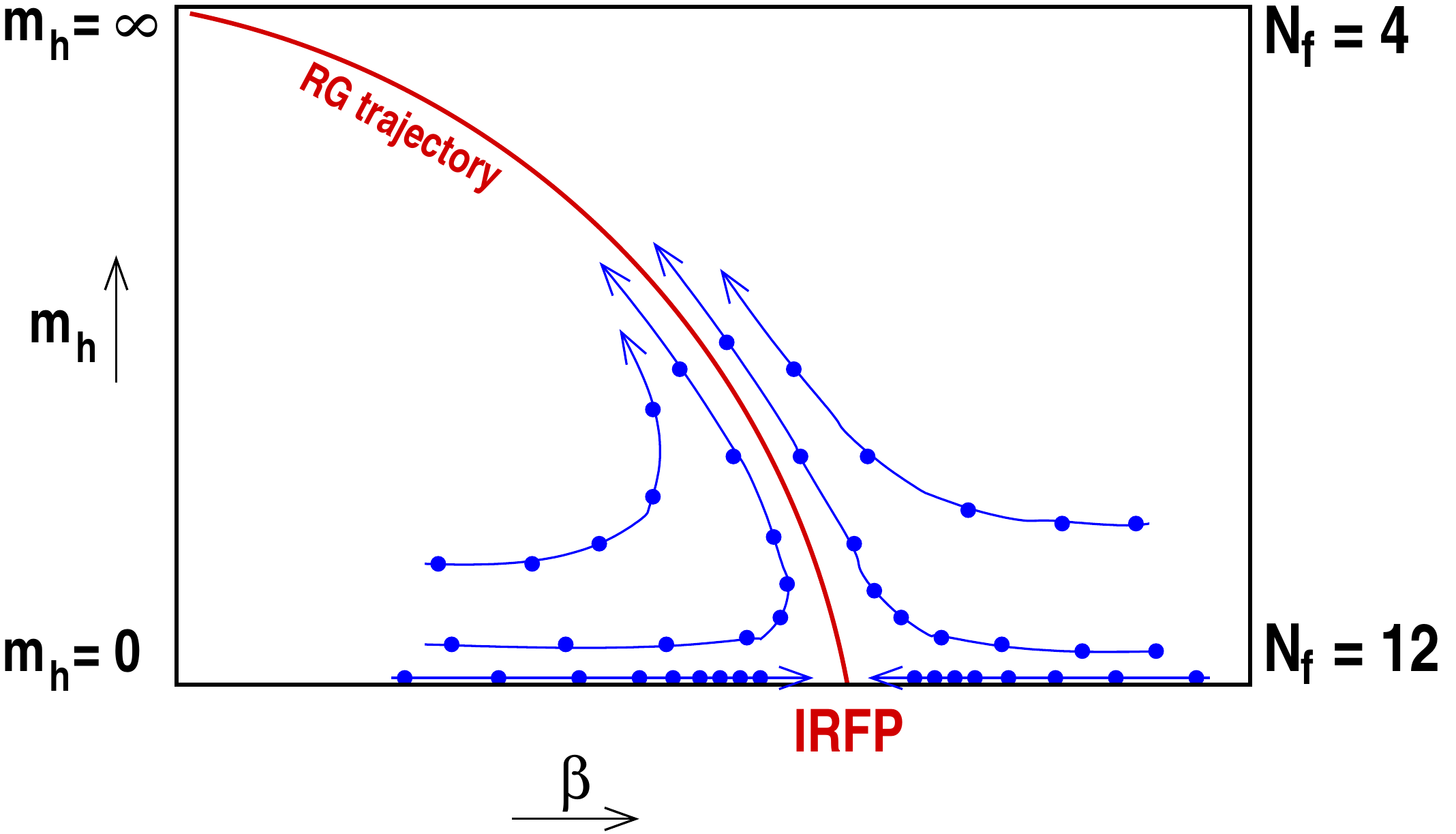}
\end{center}
    \caption{Illustration of the expected renormalization group flow lines  for the $N_\ell+N_h$ flavor theory.  The red line shows the RG trajectory connecting the conformal IRFP at $m_h=m_\ell=0$ (12 flavors) and the trivial fixed point of the 4-flavor theory at $m_h=\infty$. The blue lines are RG flow lines that first approach, then run along  the RG trajectory.  As $m_h \to 0$ the flow lines spend increasingly more time around the IRFP, creating a ``walking'' scenario, while as $m_h$ increases the heavy flavors decouple and the RG flows resemble the running of the  4-flavor system. }  
    \label{f3}
\end{figure}

In Figure \ref{f3} we sketch the renormalization group (RG) flow of the $N_\ell+N_h$ model in the limit $m_\ell=0$. In this system the mass $m_h$ of the heavy fermion is a relevant parameter. If $m_h>0$ the RG transformation traces out a trajectory that runs from the IRFP of the massless 12-flavor theory  towards the trivial fixed point of the massless 4-flavor theory. For large $m_h$  the heavy fermions decouple and the system resembles the 4-flavor chirally broken theory, while in the limit of vanishing  $m_h = m_\ell = 0$ it is conformal with the RG flow running into the IRFP. By tuning the mass of the heavy fermions we can control the  RG flow that first  approaches the IRFP, stays there for a while, and eventually runs toward the trivial fixed point. In this setup we  control the ``length'' of walking by tuning the heavy mass $m_h$, interpolating between the 12-flavor mass-deformed conformal  system in the ultraviolet and the 4-flavor chirally broken one in the infrared. 

We chose to simulate a theory with 4+8 flavors because this is the natural choice when using staggered fermions. This system  might not be the most promising BSM candidate. Simulating, e.g., a theory with $2+1$ flavors of sextet fermions could be phenomenologically more interesting since that system would have only three massless Goldstone bosons as suggested by electroweak symmetry breaking. Also, there is  increasing evidence that the $SU(3)$ 12-flavor system has a relatively small $\gamma_m^\star \approx 0.24$ anomalous dimension~\cite{Cheng:2013eu,Cheng:2013xha,Lombardo:2014pda} which may be insufficient to satisfy phenomenological walking constraints. A system with a larger anomalous dimension  might be more realistic.  However using 4+8 flavor staggered fermions allows us to embed this project into a larger program studying many flavor $SU(3)$ theories  based on simulations with the same gauge and fermion action~\cite{Hasenfratz:2011xn,Cheng:2013eu,Cheng:2011ic,Cheng:2013xha,Schaich:2013eba,Cheng:2014jba,Hasenfratz:2014rna}.  

In the following Section we will first provide details on our numerical simulations and briefly discuss properties of the generated ensembles. In Section \ref{SecRunCoupling} we present our preliminary results for determining the running coupling. As expected we observe a walking behavior in dependence of the heavy mass $m_h$. Aiming to explore the effect of a walking coupling on the ``meson spectrum'' of our theory, we show our preliminary findings based on connected and disconnected measurements in Section \ref{SecMesSpectrum} before we finally conclude.

\section{Numerical setup}
\label{SecNumSetup}

We perform our simulations using the combination of nHYP \cite{Hasenfratz:2007rf} smeared staggered fermions and the plaquette gauge action with fundamental and adjoint terms \cite{Cheng:2013xha,Cheng:2013bca}. This action with various numbers of fermion flavors has been used in numerous other works (see e.g.~\cite{Cheng:2013xha,Cheng:2013bca}). The nHYP smeared fermions have small taste breaking and simulations with our choice of smearing parameters are  numerically stable. Our ensembles of gauge field configurations are generated using the hybrid Monte Carlo (HMC) update algorithm \cite{Duane:1987de} as implemented in the FUEL software package \cite{FUEL}. We contributed our own measurement code for the connected and disconnected meson spectrum to this software project.

After an initial study with two values for the gauge coupling $\beta$, we fixed $\beta=4.0$ ($\beta_a = -\beta/4$ \cite{Hasenbusch:2004yq}) to perform our first study of the running coupling and the meson spectrum as a function of the heavy quark mass $m_h$. We present results obtained at two different lattice volumes, $L^3\times T = 24^3\times 48$ and $32^3\times 64$, with three different values for the heavy quark $m_h$ and up to five different values for the light quark $m_\ell$. We summarize our currently available configurations in Table \ref{Tab.Lattices}. Work is however in progress and we are extending some of the existing ensembles as well as increasing the number of measurements, in addition to planning new simulations at different mass values.

\begin{figure}[tb]
{\begin{picture}(149, 57)
   \put(0,1){\includegraphics[width=0.49\textwidth]{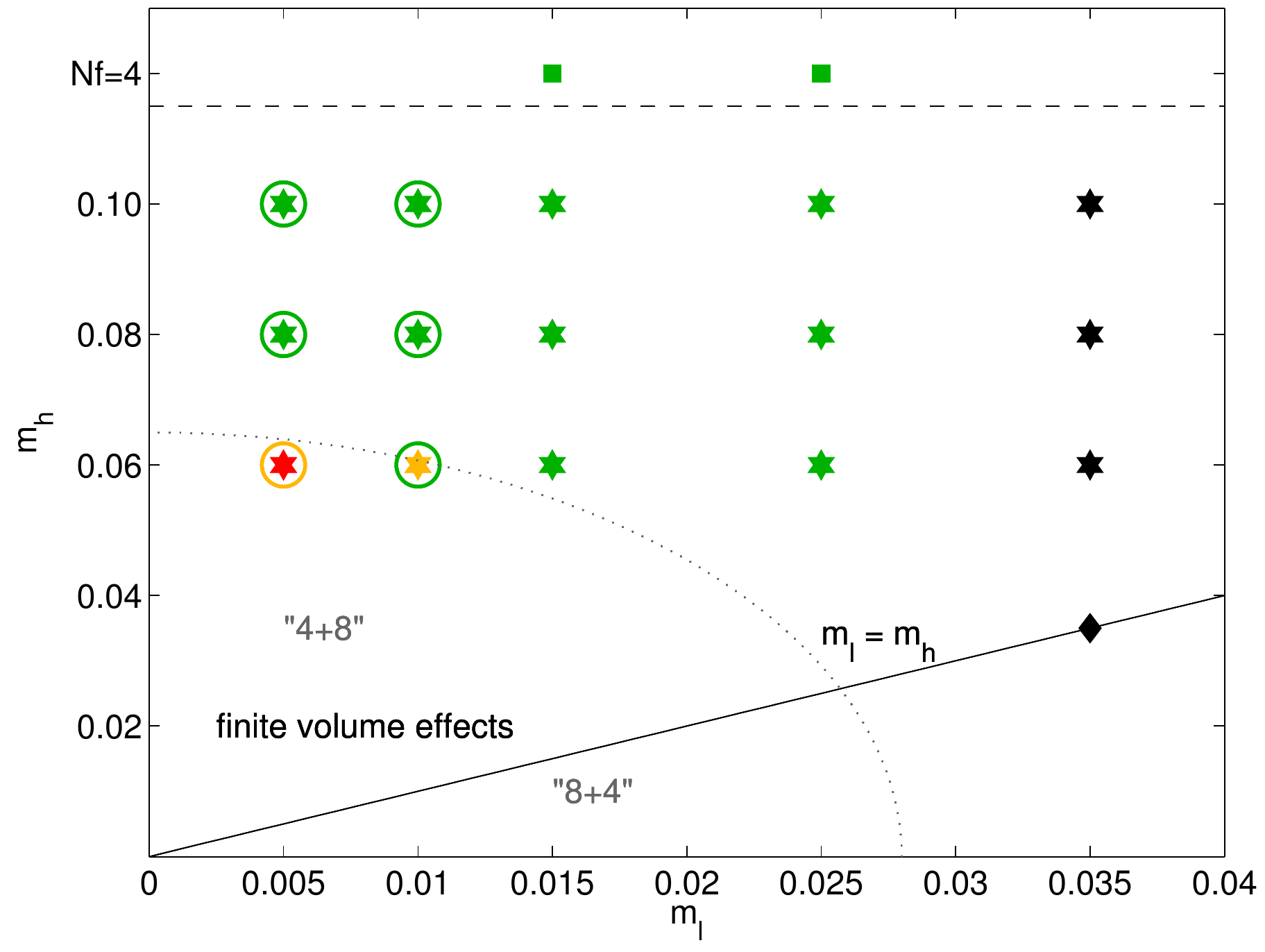}}
   \put(76,0){ \includegraphics[width=0.48\textwidth]{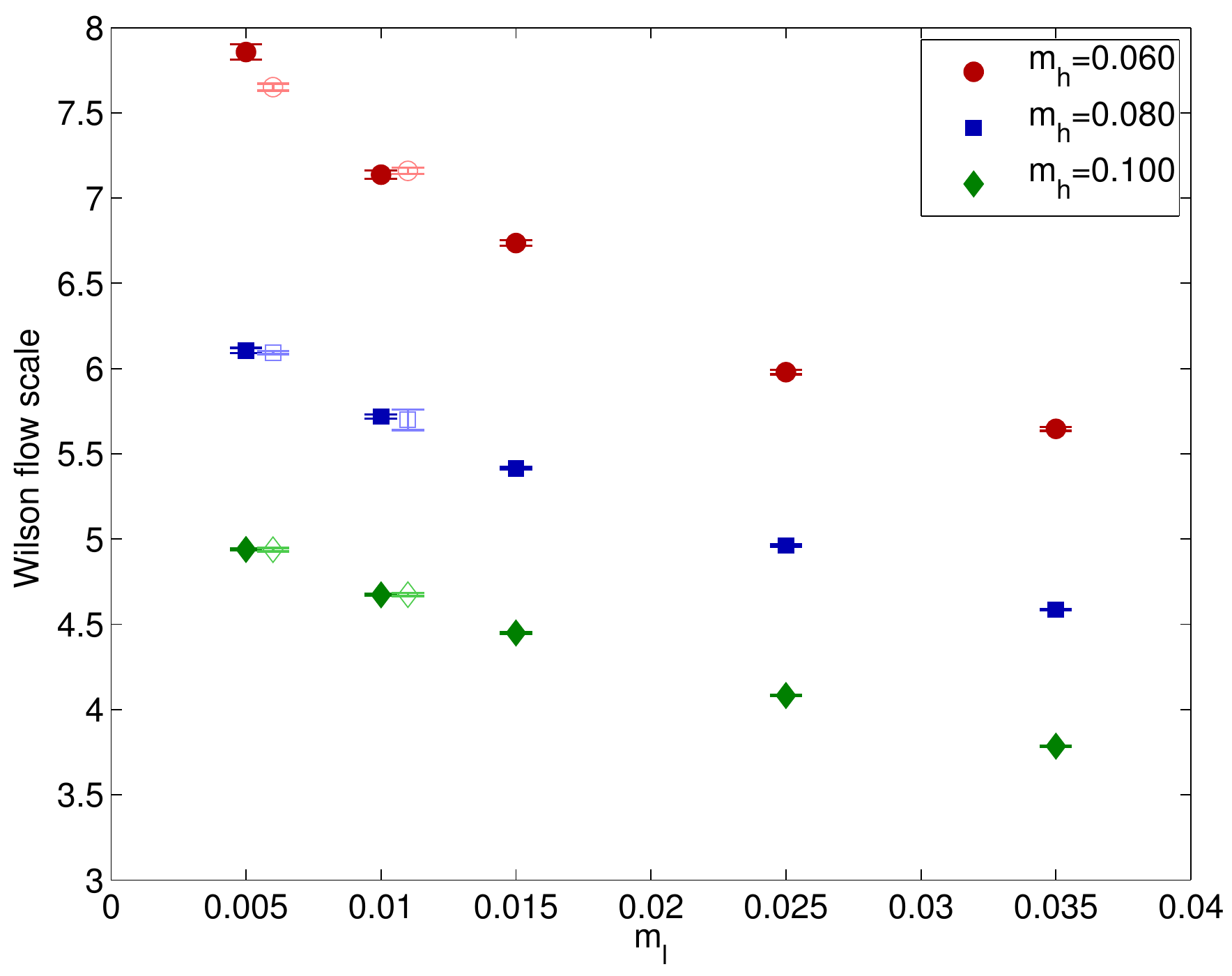}}
   \put(77,39){\rotatebox{90}{\tiny$\mathsf{\sqrt{8\tilde t_0}}$}}
\end{picture}}
    \caption{Left panel: light ($m_\ell$) and heavy ($m_h$) mass values for the simulations carried out on $24^3\times 48$ lattices (filled symbols) and $32^3\times 64$ lattice (open circles). The colors are meant to caution about finite size effects, likely negligible for green, but of increasing importance as the color turns to orange and red. The black data points are likely too heavy and affected by cut-off effects.
Right panel:  The gradient flow scale $\sqrt{8\tilde{t_0}}$ for our 21 ensembles vs.~$m_\ell$. Filled symbols show values determined on $24^3\times 48$ lattices, open symbols (shown with a small horizontal offset) refer to the values measured on $32^3\times 64$ lattices. The strong dependence on both the heavy and light fermion masses is most likely the effect of the IRFP of the 12 flavor system.}
    \label{f2}
\end{figure}

\begin{table}[tb]
\centering
\caption{Overview of our currently available ensembles at $\beta=4.0$ for up to five different values of the light quark mass $m_\ell$ and three different values for the heavy quark mass $m_h$. The table names the number of thermalized configurations, the HMC acceptance  rate, and the Wilson flow scale $\sqrt{8 \tilde t_0}$ for each ensemble. Configurations are saved every 10 trajectories and 1 trajectory = 1 MDTU. $^{\scriptscriptstyle{\heartsuit}}$ indicates HMC runs currently in progress; $^{\scriptscriptstyle{\clubsuit}}$ marks ensembles with ongoing measurements.}
\label{Tab.Lattices}
\begin{tabular}{c@{~~~}c@{~~~}r@{~~~}c@{~~~}l@{~~~}r@{~~~}c@{~~~}l@{~~~}r@{~~~}c@{~~~}l}
\toprule
              &       & \multicolumn{3}{c}{$m_h = 0.060$}& \multicolumn{3}{c}{$m_h = 0.080$}& \multicolumn{3}{c}{$m_h = 0.100$} \\              
$L^3\times T$ & $m_l$ & $N_\text{config}$ & Acc. & $\sqrt{8\tilde t_0}$ & $N_\text{config}$ & Acc.& $\sqrt{8\tilde t_0}$& $N_\text{config}$ & Acc.& $\sqrt{8\tilde t_0}$ \\ 
\midrule
$\!24^3\times 48$ & 0.005 & 1000 & 85.8\%& 7.86(4)&1000& 85.4\%& 6.11(1)&1000 &85.6\%&4.939(6)\\
$\!24^3\times 48$ & 0.010 & 1000 & 85.3\%& 7.14(3)&1000& 85.9\%& 5.72(1)&1000 &86.1\%&4.673(6)\\
$\!24^3\times 48$ & 0.015 & 1026 & 93.1\%& 6.74(2)&1000& 86.1\%& 5.415(8)&1000 &85.7\%&4.449(5)\\
$\!24^3\times 48$ & 0.025 & 1065 & 93.2\%& 5.98(1)&1020& 85.1\%& 4.962(6)&1010 &85.7\%&4.082(4)\\
$\!24^3\times 48$ & 0.035 & 1053 & 93.1\%& 5.65(1)&1020& 85.1\%& 4.587(4)&1026 &84.2\%&3.785(3)\\
\midrule
$\!32^3\times 64$ & 0.005 & 1015$^{\scriptscriptstyle{\clubsuit}}$ & 76.8\%& 7.65(2)&841$^{\scriptscriptstyle{\heartsuit\clubsuit}}$& 85.0\%& 6.093(9)&140$^{\scriptscriptstyle{\heartsuit\clubsuit}}$ &79.8\%&4.94(1)\\
$\!32^3\times 64$ & 0.010 & 1009$^{\scriptscriptstyle{\clubsuit}}$ & 77.0\%& 7.16(2)&1010$^{\scriptscriptstyle{\clubsuit}}$& 89.3\%&5.70(6) &140$^{\scriptscriptstyle{\heartsuit\clubsuit}}$&89.9\%&4.674(8)\\
\bottomrule
\end{tabular}
\end{table}
In total this work utilizes 21 ensembles which we present in a graphical overview in Fig.~\ref{f2}. The plot on the left visualizes our $24^3\times 48$ and $32^3\times 64$ ensembles in the parameter-space spanned by $m_\ell$ and $m_h$. There we also try to characterize the quality of the ensemble with respect to finite size effects using a color coding from green (no visible effects), over orange to red (severe finite size effects). Please note however this color coding is somewhat ad hoc and may need to be revisited in the future. The right plot shows the t-shift improved gradient flow scale $\sqrt{8\tilde{t_0}}$  vs.~$m_\ell$ for our 21 ensembles. (Details on the gradient flow and the  definition of the t-shift improved scale $\sqrt{8\tilde{t_0}}$ are given in the following section.) Chirally extrapolating in the light quark mass $m_\ell$, i.e.~focusing on data with the same $m_h$ (same symbol/color), we observe a non-linear increase of $\sqrt{8\tilde{t_0}}$ as $m_\ell$ decreases.  Moreover, we find a strong dependence on $m_h$ resulting in a larger scale $\sqrt{8\tilde{t_0}}$ with decreasing $m_h$. How  finite volume affects  $\sqrt{8\tilde{t_0}}$ can  be seen by comparing filled ($24^3\times 48$ data) and open symbols ($32^3\times 64$ data) with each other. For better visibility, $32^3\times 64$ data are shown with a small horizontal offset. In case of the gradient flow scale $\sqrt{8\tilde{t_0}}$ significant finite size effects are only present for our lightest data point with $m_\ell = 0.005$ and $m_h = 0.060$.

The strong dependence of the gradient flow scale on our input parameters $m_\ell$ and $m_h$ may give rise to concerns about the quality of our ensembles in particular for lighter masses $m_\ell$ and $m_h$. In order to obtain one measure on the quality, we study the evolution of the topological charge as a function of the Molecular Dynamics time $\tau$. We measure the topological charge on configurations smoothed by gradient flow transformations at flow time $t=18.0$ using an $O(a^4)$-improved definition of the topological charge 
\begin{align}
Q &= \frac{g^2}{32\pi^2}\sum_x \varepsilon_{\mu\nu\rho\sigma}\mbox{Tr}\left\{F_{\mu\nu}(x)F_{\rho\sigma}(x)\right\}, 
\end{align}
where $F_{\mu\nu}(x)$ is an appropriate linear combination of $1\times 1$, $2 \times 2$, and $3 \times 3$ clover-leaf Wilson loops as defined in~\cite{BilsonThompson:2002jk}.

\begin{figure}[tb]
\centering
\includegraphics[width=0.6\textwidth]{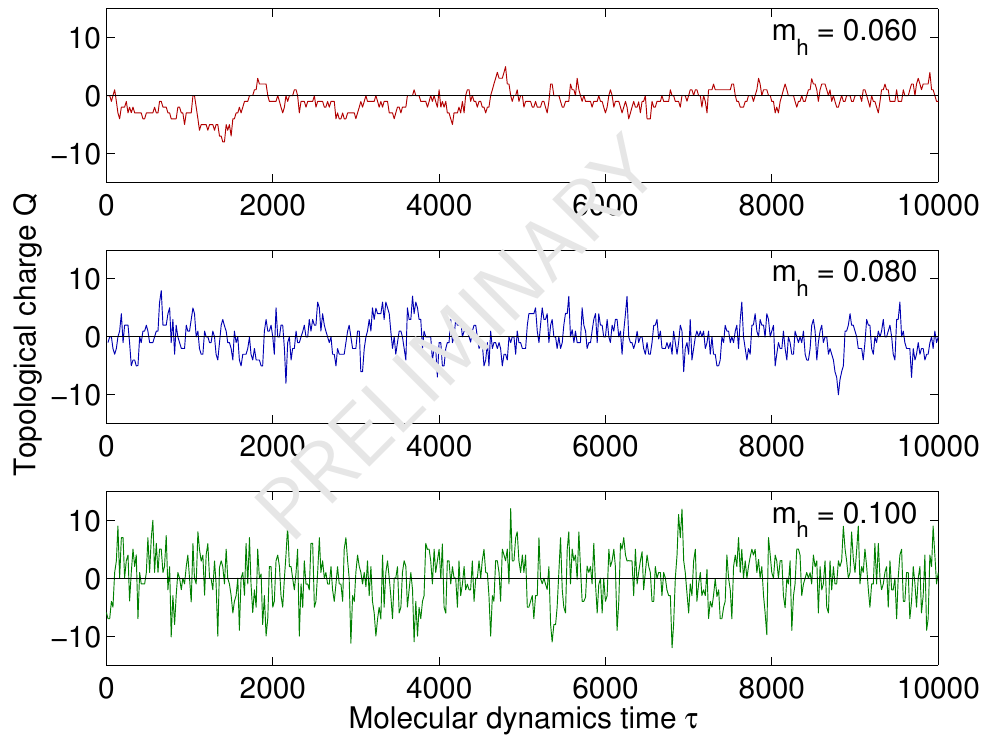}
    \caption{Topological charge as function of the Molecular Dynamics time $\tau$ for our three $24^3\times 48$ ensembles with $m_\ell=0.010$ and $m_h=0.060$, 0.080, 0.100. As we reduce the heavy mass $m_h$, slower tunneling of the topological charge and a smaller range of fluctuations occur.}
    \label{Fig.TopoCharge}
\end{figure}

 In all cases we find that the topological charge is tunneling well; we are sampling different topological sectors with an average net topological charge of near zero. As expected, by going to smaller values of $m_\ell$ and $m_h$ we do observe slower tunneling and the topological charge fluctuates with a smaller amplitude. Both leads to an increase of the integrated autocorrelation time. As an example we show in Fig.~\ref{Fig.TopoCharge} plots for the evolution of the topological charge on our $24^3$ ensembles with $m_\ell = 0.010$ and $m_h = 0.060$, 0.080, 0.100.

\section{\label{SecRunCoupling} Running Coupling}
\subsection{Scale setting using Wilson flow}

The gradient flow is an invertible and continuous smearing transformation of the gauge field. It systematically removes cut-off effects and  can therefore be used to define renormalized quantities, like the gradient flow renormalized running coupling ~\cite{Narayanan:2006rf,Luscher:2009eq,Luscher:2010iy}
\begin{equation}
g^2_{GF}(\mu) = \frac{1}{\mathcal{N}} \langle t^2 E(t) \rangle .
\label{eqn8}
\end{equation}
The flow time  $t=a^2 t_{\rm{lat}}$,  $t_{\rm{lat}}\gg 1$,  defines the energy scale $\mu^{-1} = \sqrt{8t}$ and  the energy density 
\begin{equation}
E(t) = -\frac{1}{2}{\rm Re Tr}[G_{\mu\nu}(t)G^{\mu\nu}(t)]
\label{eq7}
\end{equation}
 can be evaluated by any appropriate lattice operator. 
The constant $\mathcal{N}= 3(N^2-1) / 128\pi^2$ is chosen such that $g^2_{GF}$ matches the 
traditional $\overline{MS}$ coupling in perturbation theory~\cite{Luscher:2010iy}.

By fixing the value of the running coupling one can define a lattice scale $t_c$  
\begin{equation}
\label{eq:g2_scale}
g^2_{GF}(t_{c}) = \frac{c}{\mathcal{N}}.
\end{equation}
 In this work we use the $t_0$ scale introduced in \cite{Luscher:2010iy}
which corresponds to  $c=0.3$. There is the freedom to chose $c$ differently and we will come back to that later.

At finite lattice spacing \gGF has cut-off corrections that, for staggered fermions, are expected to be $\mathcal{O}(a^2)$
\begin{equation}
  \label{eq:lat_g2}
  \gGF(\mu; a) = \gGF(\mu; a = 0) + a^2 \mathcal{C} + \mathcal{O}(a^4 [\log a]^n,a^4).
\end{equation}
The term $a^2 \mathcal{C}$ depends on the action, the flow transformation, and the operator used to evaluate $E(t)$ in Eq.~(\ref{eq7}). It  can be significant on coarse lattices.
  Reference \cite{Cheng:2014jba} suggested a simple, empirical method to largely reduce cut-off corrections by replacing $\gGF(\mu; a)$ with 
\begin{equation}
  \label{eq:t-shift}
 \gtGF(\mu; a) = \frac{1}{\mathcal{N}} \langle t^2 E(t + \tau_0 a^2) \rangle  ,
\end{equation}
where $\tau_0 \ll t / a^2$ is a small finite shift in the flow time.
In the continuum limit $\tau_0 a^2 \to 0$ and $\gtGF(\mu) = \gGF(\mu)$.
At finite lattice spacing it is possible to choose $\tau_0$ such that  the $a^2 \mathcal{C}$ term in Eq.~(\ref{eq:lat_g2}) is canceled and
\begin{equation}
  \gtopt(\mu; a) = \gGF(\mu; a = 0) + \mathcal{O}(a^4 [\log a]^n,a^4).
\end{equation}
For full $\mathcal{O}(a^2)$ improvement the t-shift $\tau_0$  must depend on both the bare and renormalized coupling. In practice it is sufficient to choose $\tau_0$ to be a constant or only weakly  dependent  on $\gtGF(\mu)$ to remove most $\mathcal{O}(a^2)$ lattice artifacts.  Refs. \cite{AnnaTBP,AnnaLat14Talk} demonstrated that the t-shift improvement  with constant $\tau_0$ removes most lattice artifacts of the $t_0$ scale  when using the 2+1+1 flavor HISQ action.  

The reduction of lattice artifacts  is particularly important in the present work as with only one gauge coupling we are not able to take a proper continuum limit. 
To illustrate the  cut-off effects we show the unshifted $\gGF(t)$ couplings of the $m_h=0.060$, 0.080 and 0.100 systems   in   the left panel of Fig.~\ref{fig:tau_opt}.  
We use data from our $32^3\times64$ volumes with the light fermion mass extrapolated to the chiral limit $m_\ell \to 0$, and rescale  the gradient flow time $t$  by the scale factor  $t_0$.
 The rapid rise of $\gGF(t)$ at small $t$ is due to the initial integration of the gradient flow and can be considered an UV cut-off effect. 
 Beyond this initial rise, in the intermediate  energy regime the gradient flow coupling is influenced by the IRFP of the $N_f=12$ flavor system and can be different for different $m_h$ values. 
As the energy scale decreases further the heavy flavors decouple, and the running couplings of all three systems follow that of the $N_f=4$ model. 
Assuming the $t_0$ scale is already in the IR regime, we expect that the renormalized couplings are $t$ independent when $t/t_0 \gtrsim 1$.
 This is obviously not the case on the left panel of Fig.~\ref{fig:tau_opt}, signaling the presence of cut-off effects  in \gGF. 
 
A small $\tau_0$ shift can reduce the cut-off effects.  A possible way to find the optimal $\tau_0$ parameter is to require that the relative scales of the different $m_h$ systems are independent of the specific choice of the  parameter $c$  in Eq.~(\ref{eq:g2_scale}). 
 By comparing two different values,  $c=0.3$ and $c=0.35$, we find $\tau_{\rm{opt}}\approx 0.1$ to be optimal. The coupling \gtGF with this choice becomes largely independent of $\tilde t$ as the right panel   of Fig.~\ref{fig:tau_opt} shows. 
 The small deviation  observed at large $\tilde t/\tilde t_0$ is due to finite volume effects, but apart from that the three  systems predict the same renormalized running coupling for $t/t_0 \gtrsim  0.3$. This implies the corresponding scales, denoted by $\tilde{t_c}$,  have small cut-off effects. Even though the optimal t-shift was predicted in the chiral limit, we expect the same value to be also close to optimal at finite $m_\ell$~\cite{AnnaTBP}.
We compare the lattice  scales of our different ensembles on the right panel of Fig.~\ref{f2} where we show the  t-shifted $\sqrt{8\tilde{t_0}}$  in lattice units from our $24^3\times48$ and $32^3\times 64$ volume simulations. To control finite volume effects $\sqrt{8t_0} \lesssim L/5$ is usually sufficient.  This appears to be true in our case as can be seen in  Fig.~\ref{f2} or by comparing the values in Table~\ref{Tab.Lattices}.  We also find relatively strong  dependence of $\sqrt{8\tilde{t_0}}$ on the light mass $m_\ell$  indicating that we might need  larger volumes when taking the $m_\ell\to 0$ chiral limit.   

\subsection{Defining the running coupling from Wilson flow}
\begin{figure}
\begin{multicols}{2}
        \includegraphics[width=\linewidth]{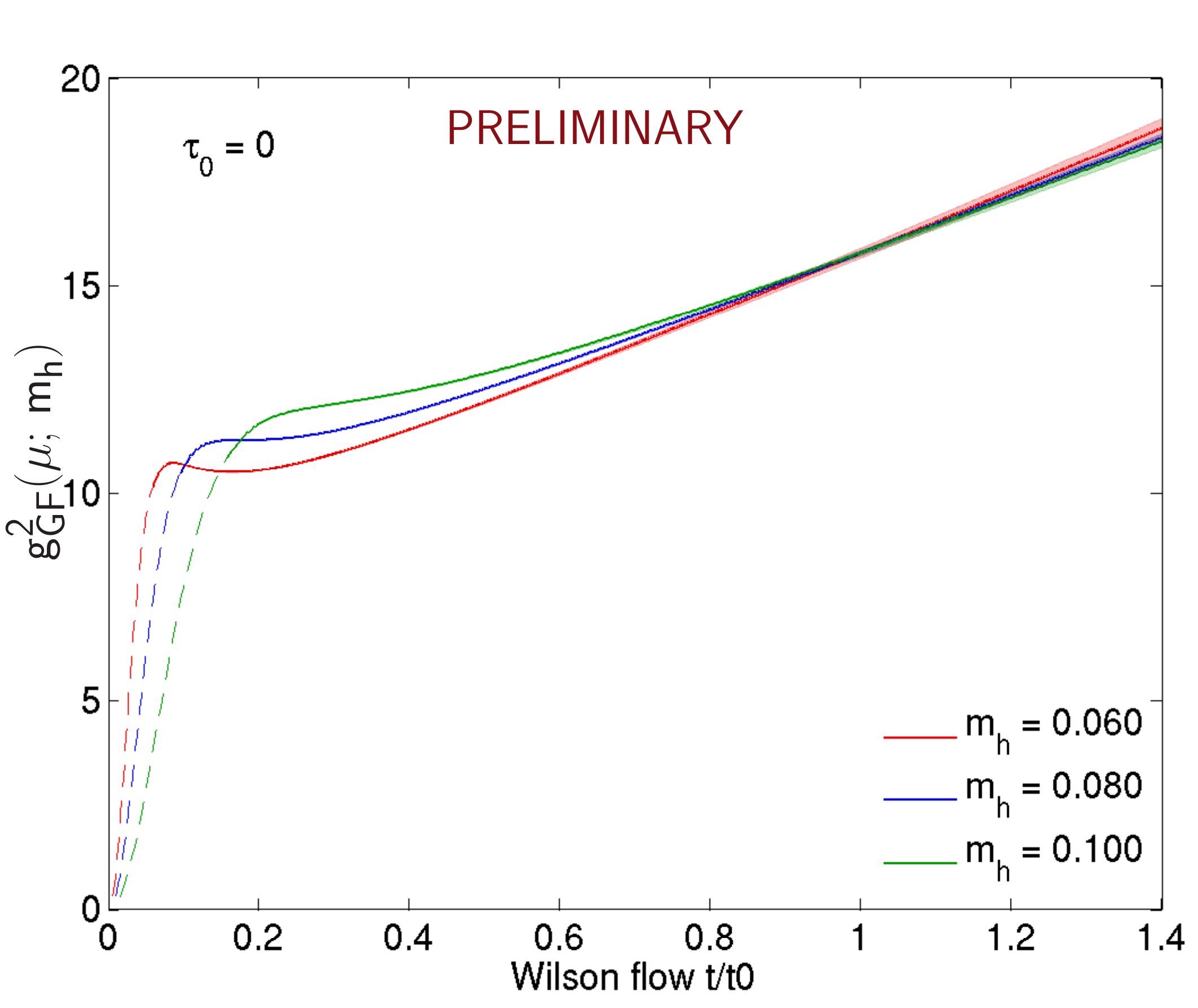}
        \includegraphics[width=\linewidth]{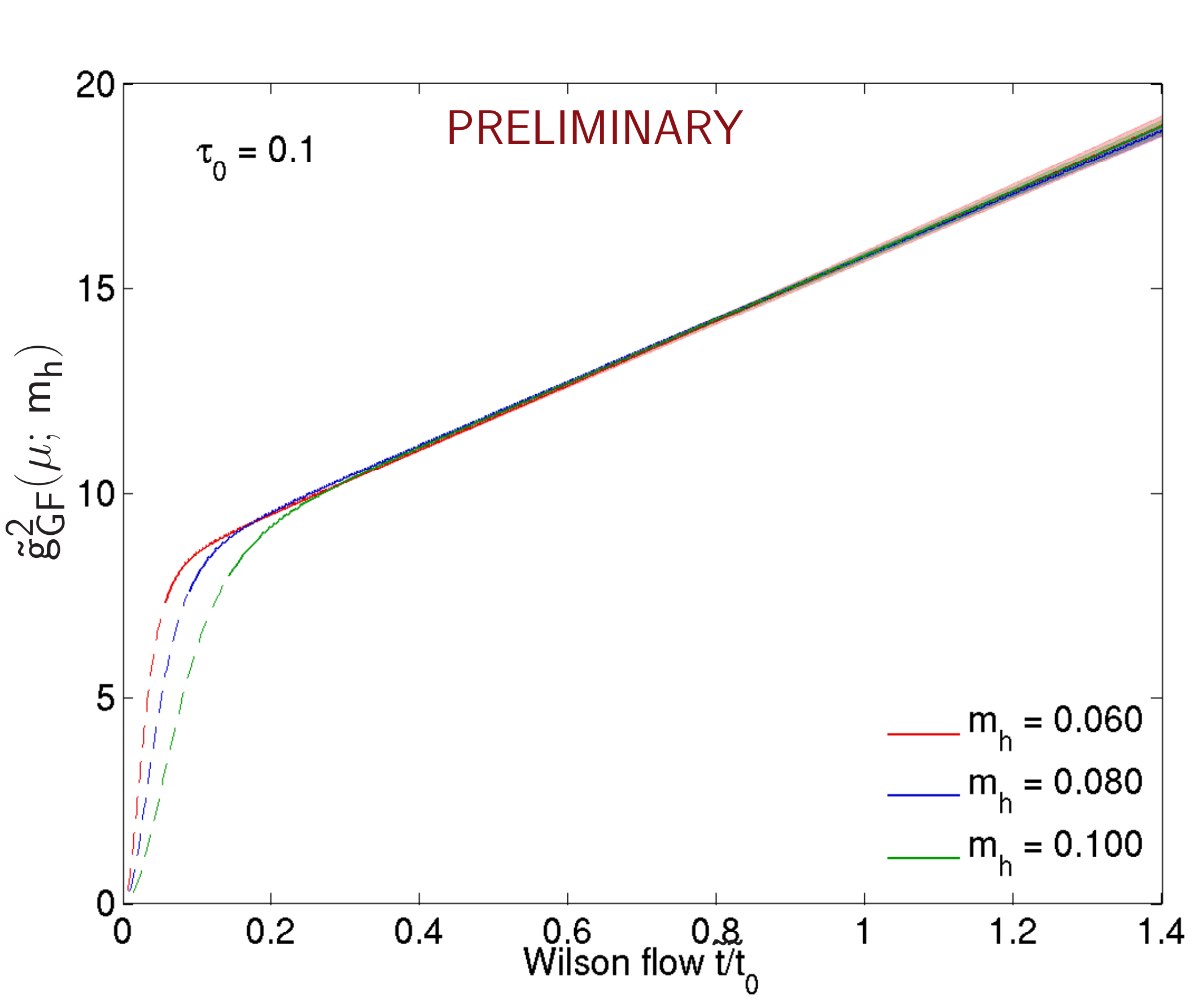}
\end{multicols}
\caption{Left panel: the gradient flow coupling $\gGF(t)$ for different values of $m_h$  
      with $m_\ell$ extrapolated to the chiral limit. The different data sets are rescaled with their corresponding $t_0$ value.  Right panel: like on the left but for the improved $\gtGF(t)$ coupling with $\tau_0=0.1$. The dashed sections of the lines indicate where we  suspect cut-off effects  may be significant.}
\label{fig:tau_opt}
\end{figure}

\begin{figure}[tb]
\begin{center}
    \includegraphics[width=8cm]{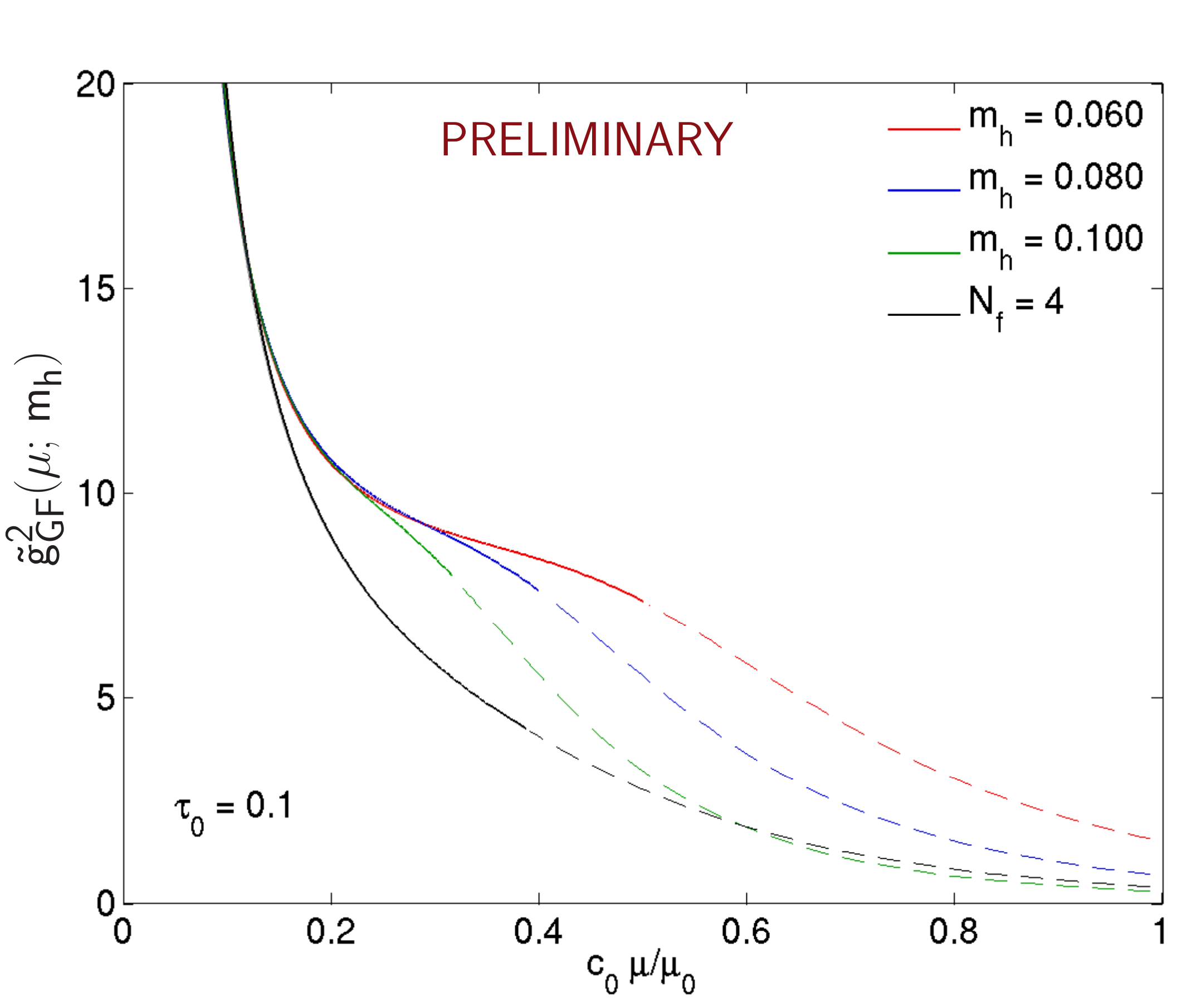} 
\end{center}
    \caption{The running coupling constant $\gtGF$ at the mass scale $\mu$ for different values of $m_h$  
      with $m_\ell$ extrapolated to the chiral limit. 
       $\mu_0$ and $c_0 = \mu_0^{-1}|_{m_h=0.060}$ serve as normalization constants that ensure that the different systems are compared at matching energy scales  and $\tau_0$ is the
      shift parameter to remove discretization errors. The dashed sections of the lines indicate where we  suspect cut-off effects  may be significant.} 
    \label{fig:running_coupling}
\end{figure}

The t-shift improved gradient flow  coupling can be  used to monitor the energy dependence of the  running coupling. Figure \ref{fig:running_coupling} shows $\gtGF(\mu)$ as the function of the energy $\mu$ for our three $N_f=4+8$ flavor ensembles and for an $N_f=4$ flavor system as well. In all cases the light masses are extrapolated to the chiral limit and we rescale $\mu$ by the lattice scale $\mu_0^{-1}=\sqrt{8\tilde{t_0}}$. Figure \ref{fig:running_coupling} is basically the same as the right panel of Fig.~\ref{fig:tau_opt} but replacing $\tilde t$ with $\mu$ reveals important features of the running coupling. At large $\mu$ the gradient flow coupling is dominated by lattice artifacts. In this region, where $\sqrt{8 \tilde{t_{\rm{lat}}}} \lesssim 2.0$,  we use dashed lines both in Fig.~\ref{fig:tau_opt} and \ref{fig:running_coupling}. In the IR limit at small $\mu$ the different systems predict a unique curve corresponding to the running coupling of the 4-flavor system. There is a clear intermediate energy region where the running coupling depends strongly on the heavy mass $m_h$. While the $N_f=4$ system shows the expected fast running, as $m_h$ decreases a shoulder develops. This is the walking behavior sought for in BSM systems. The width of the shoulder that is related to the length in energy of slow running can be tuned by tuning $m_h \to 0$, as we discussed in detail in Sec.~\ref{Intro}.  

It is important to monitor the evolution of the mass anomalous dimension. This can be done using the mode number of the Dirac operator~\cite{Cheng:2013eu,Cheng:2013bca}. Our preliminary results indicate a reasonably large anomalous dimension. The details of that calculation will be reported in a forthcoming publication.

\section{Light flavor spectrum}

\label{SecMesSpectrum}

\begin{figure}[tb]
{\begin{picture}(149, 57)
\put(0,0){\includegraphics[width=0.5\textwidth]{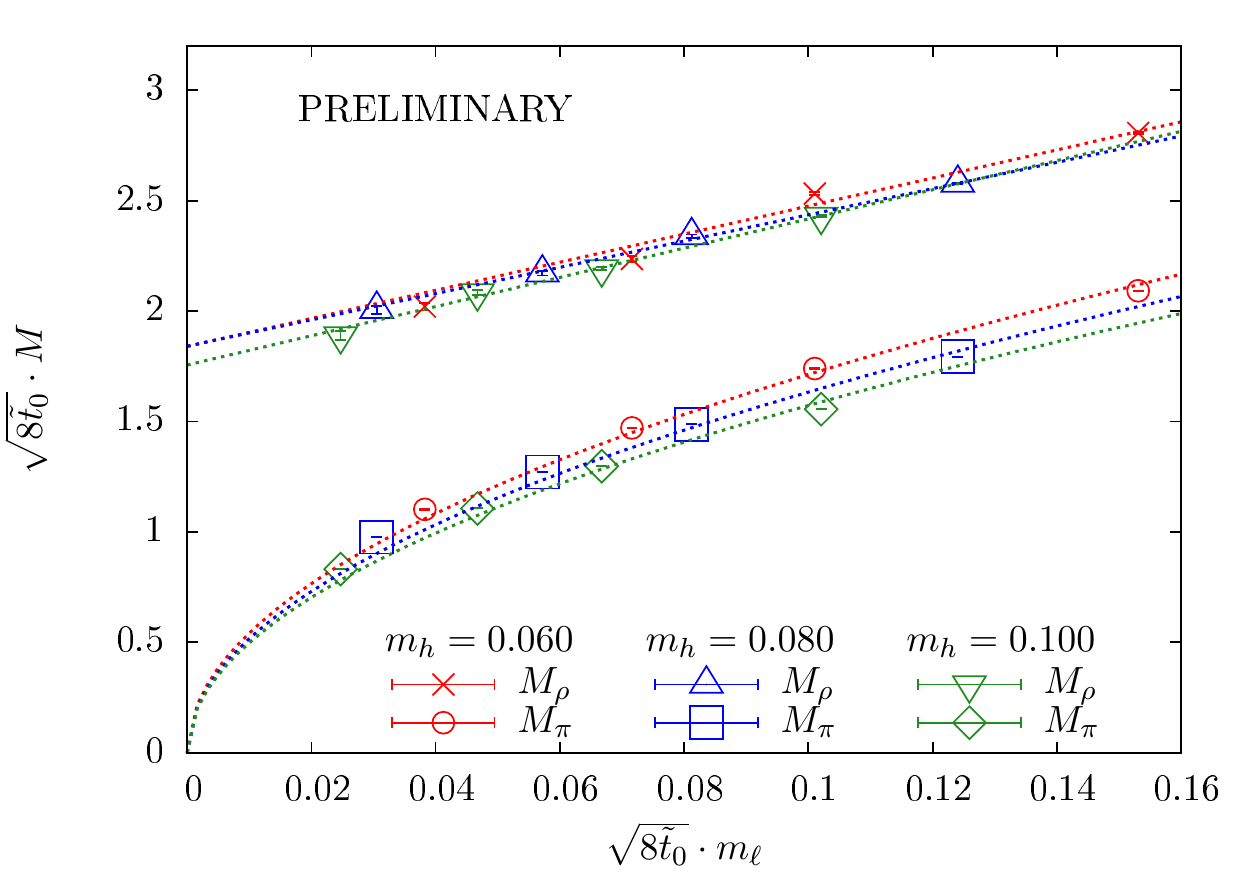}}
\put(76,0){\includegraphics[width=0.5\textwidth]{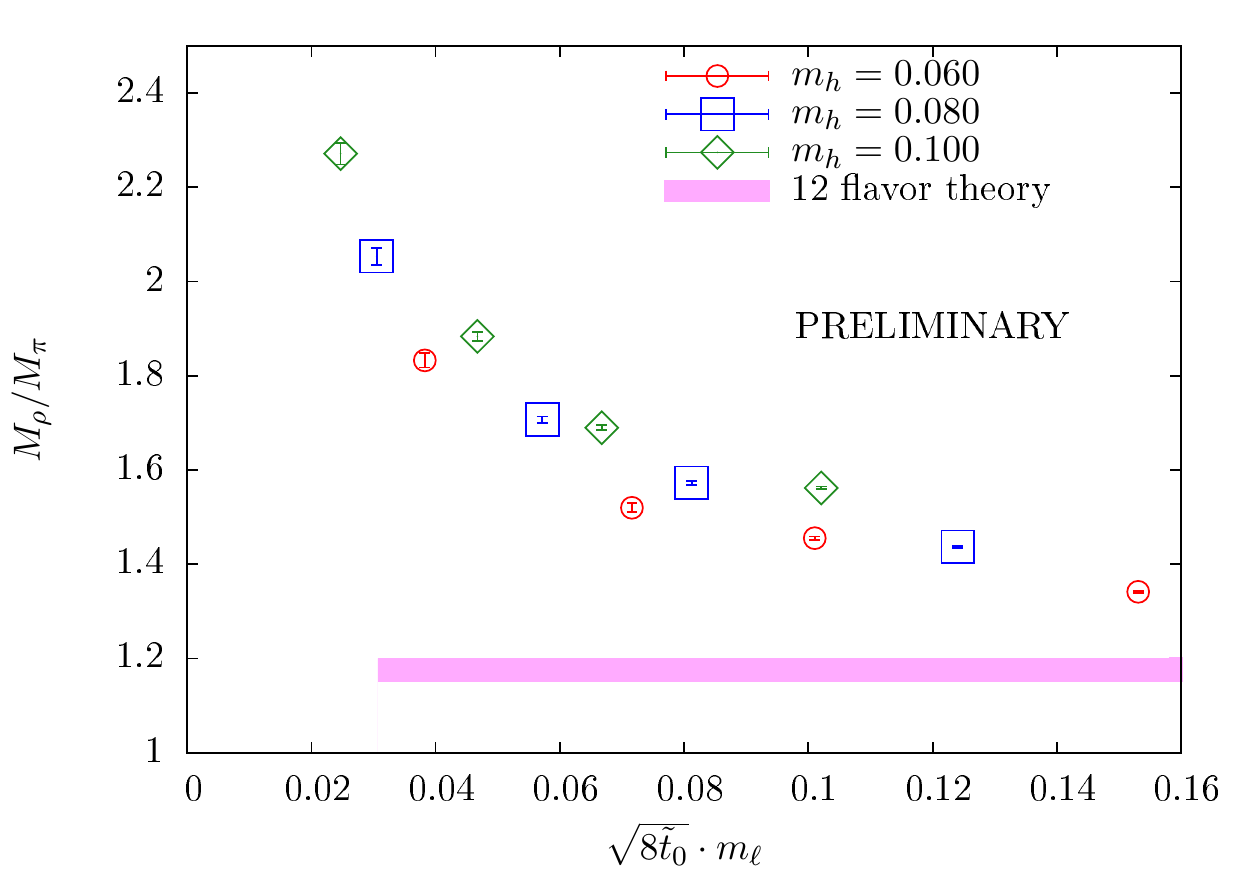}}
\end{picture}}
\caption{Left panel: the pion and rho meson masses as the function of the light fermion mass.  All masses are in units of the gradient flow scale $\sqrt{8\tilde{t_0}}$ (see Table \protect\ref{Tab.Lattices}). The dashed lines serve to guide the eye indicating the expected chiral extrapolation: linear for the masses of the rho-meson and proportional to $\sqrt{m_\ell}$ for the pion masses. Right panel: the ratio $M_\rho / M_\pi$ as the function of the light fermion mass. The shaded band around 1.17 indicates the value in the $m_h = m_\ell$ 12-flavor limit. }
\label{fig:spectrum1}
\end{figure}

In Section \ref{SecRunCoupling} we demonstrated that by tuning the mass of the heavy fermions of the $4+8$ flavor system we can control  the running  and extend the walking nature of the renormalized coupling, an important component of many BSM models.
In this section we turn our attention to the spectrum of the light flavors.  We are  especially interested in the  mass of the  $0^{++}$ iso-singlet light flavor scalar state, and its dependence on the dynamical heavy flavors. Even with our currently limited statistics,  the connected spectrum is well controlled. For the $0^{++}$ state, however, we  present only   results at a single $m_h$ value.  

A phenomenologically relevant BSM model should
predict a light iso-singlet scalar in the chiral limit, while all other non-Goldstone hadronic states  remain heavy. 
We would like to understand if such a spectrum is possible in our model and  how tuning the heavy flavors affects it.
It is useful to recall what is known about the meson spectrum of the $N_f=12$ system, the limiting case when $m_h\to m_\ell$.  In conformal systems near the chiral limit the masses of  hadronic states scale  with $\gamma_m^\star$, the anomalous dimension of the corresponding IRFP
\begin{equation}
M_H \propto m^{1/(1+\gamma_m^\star)} +\rm{corrections}.
\end{equation}
When the corrections that are  due  to the irrelevant gauge coupling and/or large fermion mass, become negligible, the ratio of any two hadronic states becomes independent of the fermion mass.  In the 12-flavor system, where the gauge coupling is nearly marginal, corrections to the scaling  can be significant. Different staggered lattice actions at various values of the gauge coupling predict the ratio of  $M_\rho / M_\pi$   between  1.0 and 1.25~\cite{Aoki:2012eq,Cheng:2013xha,Lombardo:2014pda}.  For our action at gauge coupling $\beta=4.0$  the ratio is $\approx1.17$ and due to corrections to scaling decreases slightly toward the chiral limit. 

In the $N_\ell=4$ system   chiral symmetry  is  spontaneously broken. The Goldstone boson mass scales as $M^2_\pi \propto m_\ell$, while all other hadronic states remain massive in the chiral limit. The ratio  $M_\rho/M_\pi$  diverges as $m_\ell^{-1/2}$, in sharp contrast to conformal systems.

Our results in the 4+8 flavor system for the  pion and rho are consistent with spontaneous chiral symmetry breaking. For illustration in  the left panel of Fig.~\ref{fig:spectrum1} we show the pion and rho spectrum for $m_\ell=0.005$, 0.010, 0.015 and 0.025 for all three $m_h$ values. Since the lattice scale $\tilde{t_0}$  shows strong dependence on both  the heavy and light fermion masses (see the right panel of Fig.~\ref{f2}), we rescale the lattice masses in Fig.~\ref{fig:spectrum1} by t-shift improved $\sqrt{8\tilde{t_0}}/a$ as listed in Table~\ref{Tab.Lattices}.  Results presented in this figure were obtained on the largest volume available at each mass (see Table~\ref{Tab.Lattices}). Based on the comparison of the spectrum on $24^3\times48$ and $32^3\times64$ volumes, we expect small to negligible finite volume effects for every data point with the possible exception of  $m_\ell=0.005$, $m_h=0.060$. It is  surprising how independent the rho spectrum is of the heavy fermion mass. The pion shows more variation that is further enhanced on the right panel of Fig.~\ref{fig:spectrum1} where we look at  the ratio  $M_\rho/M_\pi$   as  function of the rescaled light fermion mass. This indicates  the heavy mass influences the continuum limit even for these very basic infrared quantities. Nevertheless all three $m_h$ data sets show a rapid increase in this ratio as the light fermion mass approaches the chiral limit, as opposed to the magenta band which indicates the range  $M_\rho/M_\pi$  takes in the  12 flavor system. We do not show explicit mass dependence for $N_f=12$ as the concept of a lattice scale and  the quantity  $\sqrt{8t_0}/a$ are not well defined for a conformal system. Fortunately $M_\rho/M_\pi$ varies little with the fermion mass and the band in  Fig.~\ref{fig:spectrum1} is representative of its value.

\begin{figure}[tb]
{\begin{picture}(149, 57)
\put(0,0){\includegraphics[width=0.5\textwidth]{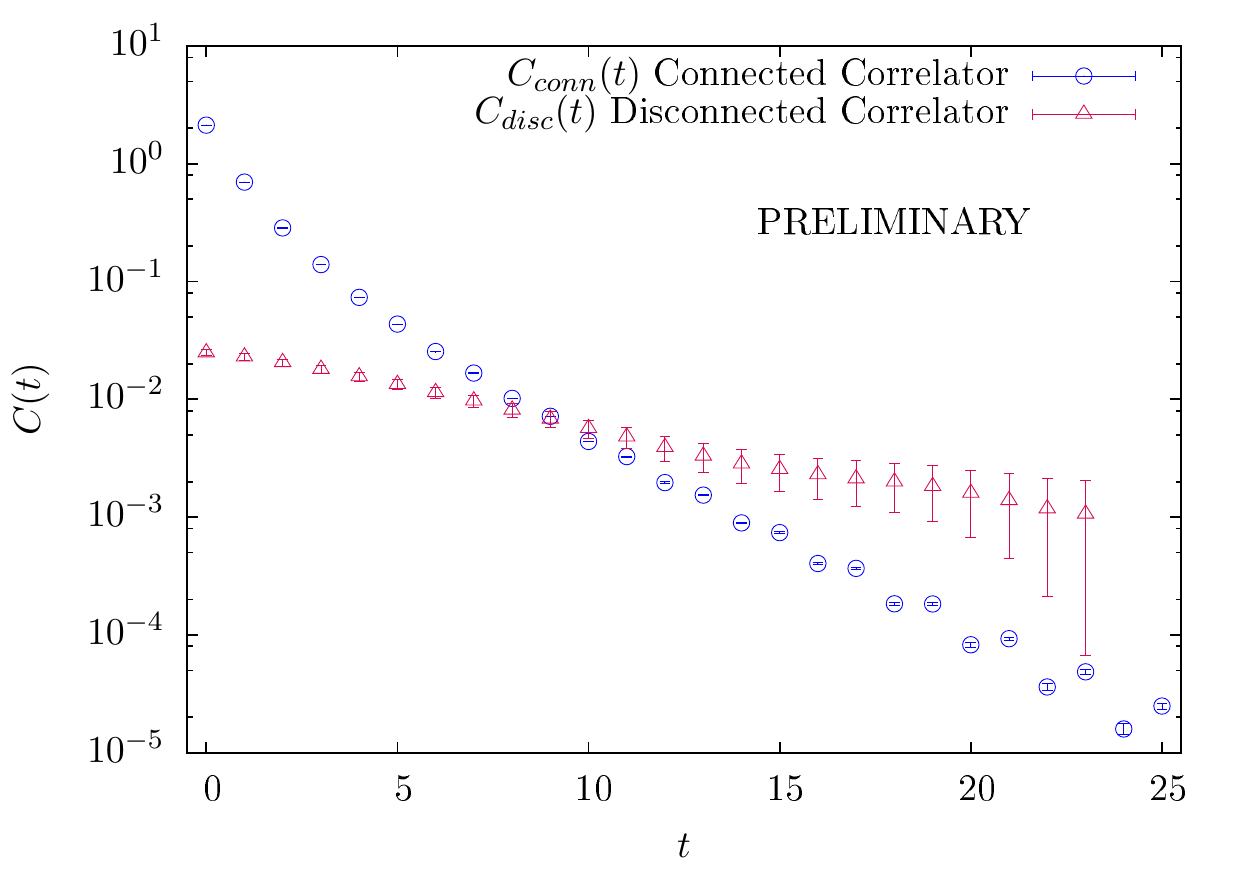}}
\put(76,0){\includegraphics[width=0.5\textwidth]{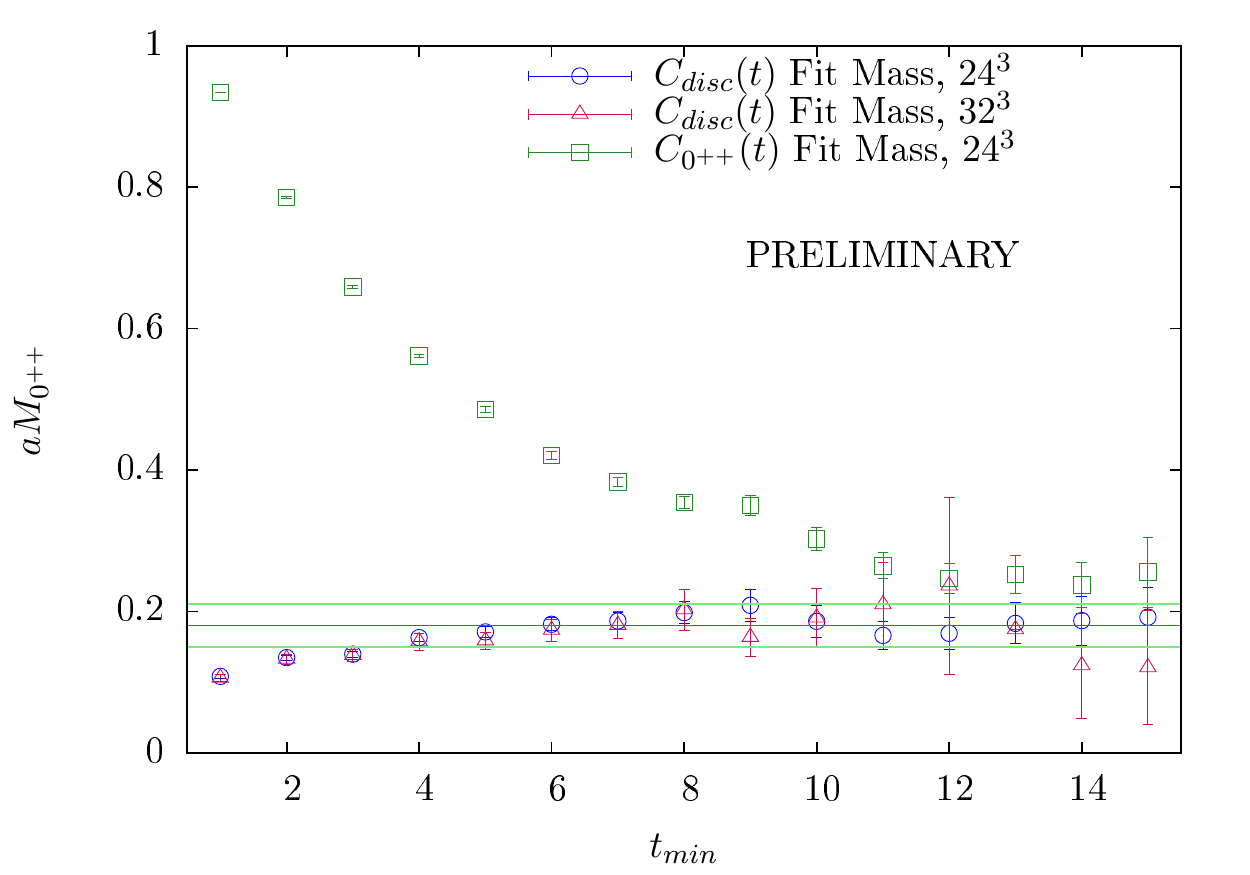}}
\end{picture}}
\caption{Left panel: Comparison of $C_{\rm{conn}}(t)$ and $C_{\rm{disc}}(t)$ for the $32^3\times 64$ $m_\ell = 0.010,\, m_h = 0.060$ ensemble, showing that $C_{\rm{conn}}(t)$ dominates the $C_{0^{++}}$ correlator for small $t$, while  for larger $t$ the noise overwhelms the $C_{\rm{disc}}(t)$ correlator. Right panel: Predictions for the $0^{++}$ mass for $m_\ell=0.010$, $m_h=0.060$.  The results are from correlated fits with a non-oscillating  plus oscillating terms in the range of $t_{\rm{min}}$ and $N_t/2$. We show results both for $24^3\times48$ and $32^3 \times64$ volumes when fitting the disconnected correlator only and $24^3\times48$ volume results when the correlator $C_{0^{++}}$ is considered.  }
\label{fig:correlator}
\end{figure}

\begin{figure}
	\begin{center}
	\includegraphics[width=0.8\linewidth]{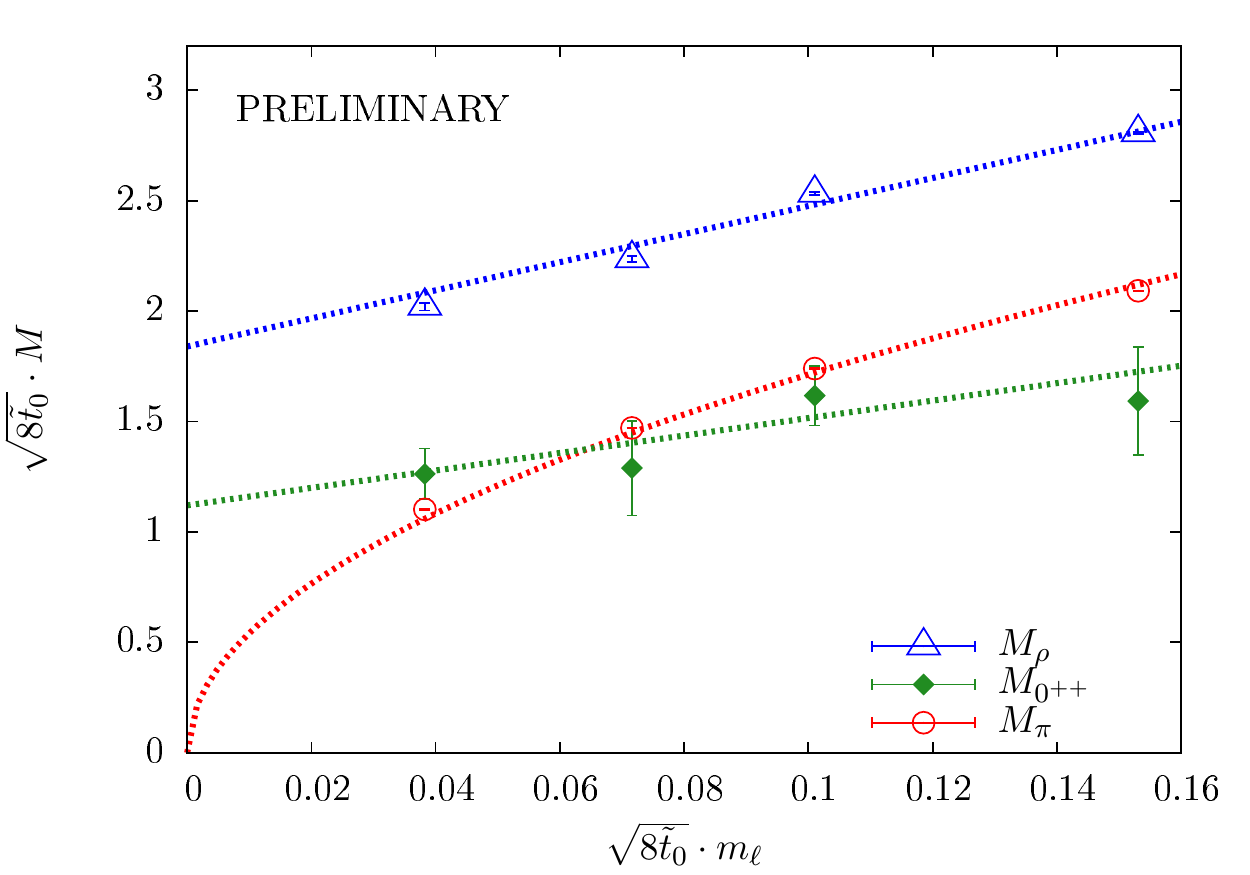}
	\end{center}
\caption{The pion, rho and $0^{++}$ scalar light flavor spectrum for $m_h=0.060$.}
\label{fig:spectrum_s}
\end{figure}

The most interesting spectral quantity, the iso-singlet $0^{++}$ $\sigma$ meson mass, is considerably more difficult to extract than the non-singlet mesons
 as it requires the evaluation of disconnected diagrams. Recent results~\cite{Aoki:2013zsa,Aoki:2014oha,Fodor:2014pqa} suggest that  in conformal and near-conformal systems the $0^{++}$ mass can be relatively light, making it somewhat easier to obtain in numerical simulations.
 The $0^{++}$ correlator is a combination of connected and disconnected diagrams
 \begin{align}
C_{0^{++}}(t) &\equiv \frac{N_\ell}{4}C_{\rm{disc} }(t) - C_{\rm{conn}}(t),
\label{eq:combined}
\end{align}
where  the  disconnected part is constructed from the vacuum subtracted operator $\langle \bar{\psi} \psi \rangle(t) - \langle\langle\bar{\psi}\psi\rangle\rangle_{\rm{e}}$, and $\langle\langle\bar{\psi}\psi\rangle\rangle_{\rm{e}}$ denotes the ensemble average of the fermion condensate. 
We construct the operator $\langle \bar{\psi} \psi \rangle(t)$ using $N_{\rm{r}}$ full volume noise sources $\eta_{i}(\vec{x},t)$ satisfying
 \begin{align}
\lim_{N_r \rightarrow \infty} \frac{1}{N_{r}} \sum_{i} \eta^\dagger_{i}(\vec{x},t) \eta_{i}(\vec{y},t') & = \delta_{\vec{x},\vec{y}} \delta_{t,t'}\, .
\end{align}
By inverting the staggered Dirac matrix
 $D((\vec{x},t), (\vec{y},t'))$  we obtain
\begin{align}
\phi_{i}(\vec{x},t) &= D((\vec{x},t),(\vec{y},t'))^{-1} \eta_{i}(\vec{y},t')
\end{align}
which leads to the scalar operator 
\begin{align}
\langle \bar{\psi} \psi \rangle(t) &= \lim_{N_r\rightarrow \infty} \frac{1}{N_r} \sum_i \sum_{\vec{x}}\eta_i^\dagger(\vec{x},t) \phi_i(\vec{x},t). 
\label{eqn:pbp}
\end{align}
On each configuration we use $N_{\rm{r}}=6$ full volume $U(1)$ noise sources diluted in time, color, as well as even/odd in space to reduce stochastic noise~\cite{Foley:2005ac}. To further enhance our signal, we use an improved operator for the chiral condensate, unique to na\"ive and staggered fermions, which replaces $\phi \eta$ in Equation (\ref{eqn:pbp}) with $m \phi \phi$. 

We compare the connected and disconnected scalar correlators on the left panel of Fig.~\ref{fig:correlator} for the $32^3\times64$, $m_\ell=0.010$, $m_h=0.060$ ensemble. At present we have evaluated the disconnected correlator on only 89	 configurations and we lose the signal in the noise for large $t$. Nevertheless the structure of the correlators is clear: For small $t$ the connected correlator dominates the combination $C_{0^{++}}$ in Eq.~(\ref{eq:combined}), the effect of the disconnected contribution does not appear until   $t\gtrsim 8$. Unfortunately by that time the signal of $C_{\rm{disc}}$ is very noisy and we are not able to identify a reliable plateau in the mass. An alternative, advocated by Refs.~\cite{Aoki:2013zsa,Aoki:2014oha,Fodor:2014pqa}, is to consider the disconnected correlator only. As long as it couples to  the $0^{++}$ state and that state is the lightest in the channel,   the iso-singlet scalar mass can be extracted from the  $C_{\rm{disc}}$ correlator alone. In practice, the disconnected correlator works even better, as apparently the excited state contributions in the connected and $0^{++}$ correlators cancel. The oscillating partner, clearly visible in the connected correlator in Fig.~\ref{fig:correlator}, is also suppressed in  $C_{\rm{disc}}$  possibly because the splitting between the iso-singlet and non-singlet partner pions is small. However, it is not absent and we observe its presence even after parity projecting the correlator. We found it more reliable to extract the $0^{++}$ mass from a correlated fit using a combination of  non-oscillating and an oscillating terms.  

The right panel of Fig.~\ref{fig:correlator} shows the $0^{++}$ mass as predicted by such correlated fits at $m_\ell=0.010$, $m_h=0.060$. The largely overlapping red and blue symbols correspond to predictions obtained on the $24^3\times48$ and $32^3\times64$ ensembles, respectively. In both cases we fit the disconnected correlator $C_{\rm{disc}}$ between time slices $t$ and $t_{\rm{max}}=N_t/2$. Predictions from the two volumes agree within errors and a plateau predicting $M_{0^{++}}=0.18(3)$ develops for $t\gtrsim 7$. We also show in the figure the predictions from the $C_{0^{++}}$ correlator of Eq.~(\ref{eq:combined}). The mass of the connected $a_0$ scalar on these ensembles is $M_{a_0}=0.360(3)$ but the connected correlator couples strongly to excited states and we do not observe a plateau even when fitting   only $C_{\rm{conn}}$  until $t\gtrsim 13$. The excited state contribution carries over to the $C_{0^{++}}$ correlator  and the predicted mass drops steadily up to $t=12$. The plateau that conceivably develops for larger $t$ is within errors consistent with the prediction obtained form the disconnected correlator only. 

Finally, in Fig.~\ref{fig:spectrum_s} we show our preliminary results for the pion, rho and $0^{++}$ scalar light flavor spectrum for the ensembles with $m_h=0.060$. As in Fig.~\ref{fig:spectrum1} we express all masses in terms of the gradient flow scale $\sqrt{8\tilde{t_0}}$. At the lightest mass we use $32^3\times64$ volumes, but even that could be volume squeezed. At all other mass values we found only small deviations between $24^3\times48$ and $32^3\times64$. Like in the  studies of conformal or near-conformal systems in Refs.~\cite{Aoki:2013zsa,Aoki:2014oha,Fodor:2014pqa}, our iso-singlet scalar is light, below the pion at the three heavier mass values. At the lightest mass the scalar state is above the pion, but the difference is not statistically significant. 
Since finite volume effects make the pion  heavy and appear to make the $0^{++}$ scalar light \cite{Aoki:2013zsa,Aoki:2014oha},  this difference could increase on larger volumes. 
In a spontaneously chirally broken system we expect the scalar mass to remain finite in the chiral limit and our data are consistent with a linear mass dependence. Our preliminary results suggests $m_\rho/m_\sigma \approx 1.7$ in the chiral limit. We want to emphasize that the results for the disconnected scalar are preliminary, based on relatively low statistics and a subset of our ensembles. However if our findings remain unchanged,  to our knowledge this is the first time that  the iso-singlet scalar mass is observed to cross the pion;  the data presented in Refs.~\cite{Aoki:2013zsa,Aoki:2014oha,Fodor:2014pqa} show the iso-singlet scalar to stay below or indistinguishable from the pion.

\section{Conclusions and outlook}
Gauge-fermion systems near the conformal window are interesting not only as strongly coupled field theory models but could also have important phenomenological applications. In recent years many large-scale lattice studies  have began to investigate systems with different gauge groups, fermion representations, and fermion numbers with the goal of identifying the onset of conformal behavior and investigating the properties of systems just below the conformal window. These investigations are, however, limited by an integer number of flavors. Our model avoids that limitation by simulating four light and eight heavy flavors where the mass of the heavy flavors is a continuous parameter allowing us to tune arbitrarily close to the conformal fixed point of the 12-flavor system in the ultraviolet, while still describing chirally broken four light flavors in the infrared limit.

We use this setup to study the energy dependence of the gauge coupling and confirm numerically a theoretically expected walking behavior. We also demonstrate that the length of walking (energy) can be changed by varying the mass of the heavy flavors. In addition, we compute the meson spectrum of the light flavors with special emphasis on the $0^{++}$ scalar. Our preliminary results reveal an iso-singlet scalar that is lighter than the pion at large fermion mass but becomes heavier at smaller masses. Further simulations will have to confirm our current findings. Measurements of the disconnected correlator are in progress on the other ensembles. While we might not be able to extract the mass of the $0^{++}$ state at larger $m_\ell$ values, we expect to be able to determine the mass of the iso-singlet scalar at least at the lightest $m_\ell$ values for each $m_h$ ensemble. It would also be very interesting to study even lighter $m_\ell$ values, possibly on larger volumes, to firmly establish the chiral behavior of the iso-singlet scalar compared to the Goldstone pion. Some of these investigations are under way and will be presented in a future publication.

\section*{Acknowledgments}
\label{acknowledgements}
Computations for this work were carried out in part on facilities of
the USQCD Collaboration, which are funded by the Office of Science of the U.S. Department of Energy, on computers at the MGHPCC, in part funded by the National Science Foundation, and on computers allocated under the NSF Xsede program to the project TG-PHY120002. \\
We thank Boston University, Fermilab, the NSF and the U.S. DOE for providing the facilities essential for the completion of this work. R.C.B., C.R. and E.W. were supported by DOE grant DE-SC0010025. In addition, R.C.B., C.R. and O.W. acknowledge the support of NSF grant OCI-0749300. A.H. acknowledges support by the DOE grant DE-SC0010005.

{\small
\bibliography{../General/BSM}

\begin{thebibliography}{35}%
\makeatletter
\providecommand \@ifxundefined [1]{%
 \@ifx{#1\undefined}
}%
\providecommand \@ifnum [1]{%
 \ifnum #1\expandafter \@firstoftwo
 \else \expandafter \@secondoftwo
 \fi
}%
\providecommand \@ifx [1]{%
 \ifx #1\expandafter \@firstoftwo
 \else \expandafter \@secondoftwo
 \fi
}%
\providecommand \natexlab [1]{#1}%
\providecommand \enquote  [1]{``#1''}%
\providecommand \bibnamefont  [1]{#1}%
\providecommand \bibfnamefont [1]{#1}%
\providecommand \citenamefont [1]{#1}%
\providecommand \href@noop [0]{\@secondoftwo}%
\providecommand \href [0]{\begingroup \@sanitize@url \@href}%
\providecommand \@href[1]{\@@startlink{#1}\@@href}%
\providecommand \@@href[1]{\endgroup#1\@@endlink}%
\providecommand \@sanitize@url [0]{\catcode `\\12\catcode `\$12\catcode
  `\&12\catcode `\#12\catcode `\^12\catcode `\_12\catcode `\%12\relax}%
\providecommand \@@startlink[1]{}%
\providecommand \@@endlink[0]{}%
\providecommand \url  [0]{\begingroup\@sanitize@url \@url }%
\providecommand \@url [1]{\endgroup\@href {#1}{\urlprefix }}%
\providecommand \urlprefix  [0]{URL }%
\providecommand \Eprint [0]{\href }%
\providecommand \doibase [0]{http://dx.doi.org/}%
\providecommand \selectlanguage [0]{\@gobble}%
\providecommand \bibinfo  [0]{\@secondoftwo}%
\providecommand \bibfield  [0]{\@secondoftwo}%
\providecommand \translation [1]{[#1]}%
\providecommand \BibitemOpen [0]{}%
\providecommand \bibitemStop [0]{}%
\providecommand \bibitemNoStop [0]{.\EOS\space}%
\providecommand \EOS [0]{\spacefactor3000\relax}%
\providecommand \BibitemShut  [1]{\csname bibitem#1\endcsname}%
\let\auto@bib@innerbib\@empty
\bibitem [{\citenamefont {Aad}\ \emph {et~al.}(2012)\citenamefont {Aad} \emph
  {et~al.}}]{Aad:2012tfa}%
  \BibitemOpen
  \bibfield  {author} {\bibinfo {author} {\bibfnamefont {G.}~\bibnamefont
  {Aad}} \emph {et~al.} (\bibinfo {collaboration} {ATLAS Collaboration}),\
  }\href {\doibase 10.1016/j.physletb.2012.08.020} {\bibfield  {journal}
  {\bibinfo  {journal} {Phys.Lett.}\ }\textbf {\bibinfo {volume} {B716}},\
  \bibinfo {pages} {1} (\bibinfo {year} {2012})},\ \Eprint
  {http://arxiv.org/abs/1207.7214} {arXiv:1207.7214 [hep-ex]} \BibitemShut
  {NoStop}%
\bibitem [{\citenamefont {Chatrchyan}\ \emph {et~al.}(2012)\citenamefont
  {Chatrchyan} \emph {et~al.}}]{Chatrchyan:2012ufa}%
  \BibitemOpen
  \bibfield  {author} {\bibinfo {author} {\bibfnamefont {S.}~\bibnamefont
  {Chatrchyan}} \emph {et~al.} (\bibinfo {collaboration} {CMS Collaboration}),\
  }\href {\doibase 10.1016/j.physletb.2012.08.021} {\bibfield  {journal}
  {\bibinfo  {journal} {Phys.Lett.}\ }\textbf {\bibinfo {volume} {B716}},\
  \bibinfo {pages} {30} (\bibinfo {year} {2012})},\ \Eprint
  {http://arxiv.org/abs/1207.7235} {arXiv:1207.7235 [hep-ex]} \BibitemShut
  {NoStop}%
\bibitem [{\citenamefont {Weinberg}(1979)}]{Weinberg:1979bn}%
  \BibitemOpen
  \bibfield  {author} {\bibinfo {author} {\bibfnamefont {S.}~\bibnamefont
  {Weinberg}},\ }\href {\doibase 10.1103/PhysRevD.19.1277} {\bibfield
  {journal} {\bibinfo  {journal} {Phys.Rev.}\ }\textbf {\bibinfo {volume}
  {D19}},\ \bibinfo {pages} {1277} (\bibinfo {year} {1979})}\BibitemShut
  {NoStop}%
\bibitem [{\citenamefont {Susskind}(1979)}]{Susskind:1978ms}%
  \BibitemOpen
  \bibfield  {author} {\bibinfo {author} {\bibfnamefont {L.}~\bibnamefont
  {Susskind}},\ }\href {\doibase 10.1103/PhysRevD.20.2619} {\bibfield
  {journal} {\bibinfo  {journal} {Phys.Rev.}\ }\textbf {\bibinfo {volume}
  {D20}},\ \bibinfo {pages} {2619} (\bibinfo {year} {1979})}\BibitemShut
  {NoStop}%
\bibitem [{\citenamefont {Dimopoulos}\ and\ \citenamefont
  {Susskind}(1979)}]{Dimopoulos:1979es}%
  \BibitemOpen
  \bibfield  {author} {\bibinfo {author} {\bibfnamefont {S.}~\bibnamefont
  {Dimopoulos}}\ and\ \bibinfo {author} {\bibfnamefont {L.}~\bibnamefont
  {Susskind}},\ }\href {\doibase 10.1016/0550-3213(79)90364-X} {\bibfield
  {journal} {\bibinfo  {journal} {Nucl.Phys.}\ }\textbf {\bibinfo {volume}
  {B155}},\ \bibinfo {pages} {237} (\bibinfo {year} {1979})}\BibitemShut
  {NoStop}%
\bibitem [{\citenamefont {Eichten}\ and\ \citenamefont
  {Lane}(1980)}]{Eichten:1979ah}%
  \BibitemOpen
  \bibfield  {author} {\bibinfo {author} {\bibfnamefont {E.}~\bibnamefont
  {Eichten}}\ and\ \bibinfo {author} {\bibfnamefont {K.~D.}\ \bibnamefont
  {Lane}},\ }\href {\doibase 10.1016/0370-2693(80)90065-9} {\bibfield
  {journal} {\bibinfo  {journal} {Phys.Lett.}\ }\textbf {\bibinfo {volume}
  {B90}},\ \bibinfo {pages} {125} (\bibinfo {year} {1980})}\BibitemShut
  {NoStop}%
\bibitem [{\citenamefont {Yamawaki}\ \emph {et~al.}(1986)\citenamefont
  {Yamawaki}, \citenamefont {Bando},\ and\ \citenamefont
  {Matumoto}}]{Yamawaki:1985zg}%
  \BibitemOpen
  \bibfield  {author} {\bibinfo {author} {\bibfnamefont {K.}~\bibnamefont
  {Yamawaki}}, \bibinfo {author} {\bibfnamefont {M.}~\bibnamefont {Bando}}, \
  and\ \bibinfo {author} {\bibfnamefont {K.-i.}\ \bibnamefont {Matumoto}},\
  }\href {\doibase 10.1103/PhysRevLett.56.1335} {\bibfield  {journal} {\bibinfo
   {journal} {Phys. Rev. Lett.}\ }\textbf {\bibinfo {volume} {56}},\ \bibinfo
  {pages} {1335} (\bibinfo {year} {1986})}\BibitemShut {NoStop}%
\bibitem [{\citenamefont {Appelquist}\ \emph {et~al.}(1991)\citenamefont
  {Appelquist}, \citenamefont {Terning},\ and\ \citenamefont
  {Wijewardhana}}]{Appelquist:1991nm}%
  \BibitemOpen
  \bibfield  {author} {\bibinfo {author} {\bibfnamefont {T.}~\bibnamefont
  {Appelquist}}, \bibinfo {author} {\bibfnamefont {J.}~\bibnamefont {Terning}},
  \ and\ \bibinfo {author} {\bibfnamefont {L.~C.~R.}\ \bibnamefont
  {Wijewardhana}},\ }\href {\doibase 10.1103/PhysRevD.44.871} {\bibfield
  {journal} {\bibinfo  {journal} {Phys. Rev. D}\ }\textbf {\bibinfo {volume}
  {44}},\ \bibinfo {pages} {871} (\bibinfo {year} {1991})}\BibitemShut
  {NoStop}%
\bibitem [{\citenamefont {Appelquist}\ \emph {et~al.}(2011)\citenamefont
  {Appelquist}, \citenamefont {Fleming}, \citenamefont {Lin}, \citenamefont
  {Neil},\ and\ \citenamefont {Schaich}}]{Appelquist:2011dp}%
  \BibitemOpen
  \bibfield  {author} {\bibinfo {author} {\bibfnamefont {T.}~\bibnamefont
  {Appelquist}}, \bibinfo {author} {\bibfnamefont {G.}~\bibnamefont {Fleming}},
  \bibinfo {author} {\bibfnamefont {M.}~\bibnamefont {Lin}}, \bibinfo {author}
  {\bibfnamefont {E.}~\bibnamefont {Neil}}, \ and\ \bibinfo {author}
  {\bibfnamefont {D.}~\bibnamefont {Schaich}},\ }\href {\doibase
  10.1103/PhysRevD.84.054501} {\bibfield  {journal} {\bibinfo  {journal}
  {Phys.Rev.}\ }\textbf {\bibinfo {volume} {D84}},\ \bibinfo {pages} {054501}
  (\bibinfo {year} {2011})},\ \Eprint {http://arxiv.org/abs/1106.2148}
  {arXiv:1106.2148 [hep-lat]} \BibitemShut {NoStop}%
\bibitem [{\citenamefont {Hasenfratz}(2012)}]{Hasenfratz:2011xn}%
  \BibitemOpen
  \bibfield  {author} {\bibinfo {author} {\bibfnamefont {A.}~\bibnamefont
  {Hasenfratz}},\ }\href {\doibase 10.1103/PhysRevLett.108.061601} {\bibfield
  {journal} {\bibinfo  {journal} {Phys.Rev.Lett.}\ }\textbf {\bibinfo {volume}
  {108}},\ \bibinfo {pages} {061601} (\bibinfo {year} {2012})},\ \Eprint
  {http://arxiv.org/abs/1106.5293} {arXiv:1106.5293 [hep-lat]} \BibitemShut
  {NoStop}%
\bibitem [{\citenamefont {Aoki}\ \emph {et~al.}(2012)\citenamefont {Aoki},
  \citenamefont {Aoyama}, \citenamefont {Kurachi}, \citenamefont {Maskawa},
  \citenamefont {Nagai} \emph {et~al.}}]{Aoki:2012eq}%
  \BibitemOpen
  \bibfield  {author} {\bibinfo {author} {\bibfnamefont {Y.}~\bibnamefont
  {Aoki}}, \bibinfo {author} {\bibfnamefont {T.}~\bibnamefont {Aoyama}},
  \bibinfo {author} {\bibfnamefont {M.}~\bibnamefont {Kurachi}}, \bibinfo
  {author} {\bibfnamefont {T.}~\bibnamefont {Maskawa}}, \bibinfo {author}
  {\bibfnamefont {K.-i.}\ \bibnamefont {Nagai}},  \emph {et~al.},\ }\href
  {\doibase 10.1103/PhysRevD.86.059903, 10.1103/PhysRevD.86.054506} {\bibfield
  {journal} {\bibinfo  {journal} {Phys.Rev.}\ }\textbf {\bibinfo {volume}
  {D86}},\ \bibinfo {pages} {054506} (\bibinfo {year} {2012})},\ \Eprint
  {http://arxiv.org/abs/1207.3060} {arXiv:1207.3060 [hep-lat]} \BibitemShut
  {NoStop}%
\bibitem [{\citenamefont {DeGrand}(2011)}]{DeGrand:2011cu}%
  \BibitemOpen
  \bibfield  {author} {\bibinfo {author} {\bibfnamefont {T.}~\bibnamefont
  {DeGrand}},\ }\href {\doibase 10.1103/PhysRevD.84.116901} {\bibfield
  {journal} {\bibinfo  {journal} {Phys.Rev.}\ }\textbf {\bibinfo {volume}
  {D84}},\ \bibinfo {pages} {116901} (\bibinfo {year} {2011})},\ \Eprint
  {http://arxiv.org/abs/1109.1237} {arXiv:1109.1237 [hep-lat]} \BibitemShut
  {NoStop}%
\bibitem [{\citenamefont {Cheng}\ \emph
  {et~al.}(2014{\natexlab{a}})\citenamefont {Cheng}, \citenamefont
  {Hasenfratz}, \citenamefont {Liu}, \citenamefont {Petropoulos},\ and\
  \citenamefont {Schaich}}]{Cheng:2013xha}%
  \BibitemOpen
  \bibfield  {author} {\bibinfo {author} {\bibfnamefont {A.}~\bibnamefont
  {Cheng}}, \bibinfo {author} {\bibfnamefont {A.}~\bibnamefont {Hasenfratz}},
  \bibinfo {author} {\bibfnamefont {Y.}~\bibnamefont {Liu}}, \bibinfo {author}
  {\bibfnamefont {G.}~\bibnamefont {Petropoulos}}, \ and\ \bibinfo {author}
  {\bibfnamefont {D.}~\bibnamefont {Schaich}},\ }\href {\doibase
  10.1103/PhysRevD.90.014509} {\bibfield  {journal} {\bibinfo  {journal}
  {Phys.Rev.}\ }\textbf {\bibinfo {volume} {D90}},\ \bibinfo {pages} {014509}
  (\bibinfo {year} {2014}{\natexlab{a}})},\ \Eprint
  {http://arxiv.org/abs/1401.0195} {arXiv:1401.0195 [hep-lat]} \BibitemShut
  {NoStop}%
\bibitem [{\citenamefont {Cheng}\ \emph
  {et~al.}(2014{\natexlab{b}})\citenamefont {Cheng}, \citenamefont
  {Hasenfratz}, \citenamefont {Liu}, \citenamefont {Petropoulos},\ and\
  \citenamefont {Schaich}}]{Cheng:2014jba}%
  \BibitemOpen
  \bibfield  {author} {\bibinfo {author} {\bibfnamefont {A.}~\bibnamefont
  {Cheng}}, \bibinfo {author} {\bibfnamefont {A.}~\bibnamefont {Hasenfratz}},
  \bibinfo {author} {\bibfnamefont {Y.}~\bibnamefont {Liu}}, \bibinfo {author}
  {\bibfnamefont {G.}~\bibnamefont {Petropoulos}}, \ and\ \bibinfo {author}
  {\bibfnamefont {D.}~\bibnamefont {Schaich}},\ }\href {\doibase
  10.1007/JHEP05(2014)137} {\bibfield  {journal} {\bibinfo  {journal} {JHEP}\
  }\textbf {\bibinfo {volume} {1405}},\ \bibinfo {pages} {137} (\bibinfo {year}
  {2014}{\natexlab{b}})},\ \Eprint {http://arxiv.org/abs/1404.0984}
  {arXiv:1404.0984 [hep-lat]} \BibitemShut {NoStop}%
\bibitem [{\citenamefont {Lombardo}\ \emph {et~al.}(2014)\citenamefont
  {Lombardo}, \citenamefont {Miura}, \citenamefont {da~Silva},\ and\
  \citenamefont {Pallante}}]{Lombardo:2014pda}%
  \BibitemOpen
  \bibfield  {author} {\bibinfo {author} {\bibfnamefont {M.}~\bibnamefont
  {Lombardo}}, \bibinfo {author} {\bibfnamefont {K.}~\bibnamefont {Miura}},
  \bibinfo {author} {\bibfnamefont {T.~J.~N.}\ \bibnamefont {da~Silva}}, \ and\
  \bibinfo {author} {\bibfnamefont {E.}~\bibnamefont {Pallante}},\ }\href@noop
  {} {\  (\bibinfo {year} {2014})},\ \Eprint {http://arxiv.org/abs/1410.0298}
  {arXiv:1410.0298 [hep-lat]} \BibitemShut {NoStop}%
\bibitem [{\citenamefont {Aoki}\ \emph {et~al.}(2013)\citenamefont {Aoki},
  \citenamefont {Aoyama}, \citenamefont {Kurachi}, \citenamefont {Maskawa},
  \citenamefont {Nagai} \emph {et~al.}}]{Aoki:2013zsa}%
  \BibitemOpen
  \bibfield  {author} {\bibinfo {author} {\bibfnamefont {Y.}~\bibnamefont
  {Aoki}}, \bibinfo {author} {\bibfnamefont {T.}~\bibnamefont {Aoyama}},
  \bibinfo {author} {\bibfnamefont {M.}~\bibnamefont {Kurachi}}, \bibinfo
  {author} {\bibfnamefont {T.}~\bibnamefont {Maskawa}}, \bibinfo {author}
  {\bibfnamefont {K.-i.}\ \bibnamefont {Nagai}},  \emph {et~al.},\ }\href
  {\doibase 10.1103/PhysRevLett.111.162001} {\bibfield  {journal} {\bibinfo
  {journal} {Phys.Rev.Lett.}\ }\textbf {\bibinfo {volume} {111}},\ \bibinfo
  {pages} {162001} (\bibinfo {year} {2013})},\ \Eprint
  {http://arxiv.org/abs/1305.6006} {arXiv:1305.6006 [hep-lat]} \BibitemShut
  {NoStop}%
\bibitem [{\citenamefont {Aoki}\ \emph {et~al.}(2014)\citenamefont {Aoki} \emph
  {et~al.}}]{Aoki:2014oha}%
  \BibitemOpen
  \bibfield  {author} {\bibinfo {author} {\bibfnamefont {Y.}~\bibnamefont
  {Aoki}} \emph {et~al.} (\bibinfo {collaboration} {the LatKMI
  Collaboration}),\ }\href {\doibase 10.1103/PhysRevD.89.111502} {\bibfield
  {journal} {\bibinfo  {journal} {Phys.Rev.}\ }\textbf {\bibinfo {volume}
  {D89}},\ \bibinfo {pages} {111502} (\bibinfo {year} {2014})},\ \Eprint
  {http://arxiv.org/abs/1403.5000} {arXiv:1403.5000 [hep-lat]} \BibitemShut
  {NoStop}%
\bibitem [{\citenamefont {Fodor}\ \emph {et~al.}(2014)\citenamefont {Fodor},
  \citenamefont {Holland}, \citenamefont {Kuti}, \citenamefont {Nogradi},\ and\
  \citenamefont {Wong}}]{Fodor:2014pqa}%
  \BibitemOpen
  \bibfield  {author} {\bibinfo {author} {\bibfnamefont {Z.}~\bibnamefont
  {Fodor}}, \bibinfo {author} {\bibfnamefont {K.}~\bibnamefont {Holland}},
  \bibinfo {author} {\bibfnamefont {J.}~\bibnamefont {Kuti}}, \bibinfo {author}
  {\bibfnamefont {D.}~\bibnamefont {Nogradi}}, \ and\ \bibinfo {author}
  {\bibfnamefont {C.~H.}\ \bibnamefont {Wong}},\ }\href@noop {} {\bibfield
  {journal} {\bibinfo  {journal} {PoS}\ }\textbf {\bibinfo {volume}
  {LATTICE2013}},\ \bibinfo {pages} {062} (\bibinfo {year} {2014})},\ \Eprint
  {http://arxiv.org/abs/1401.2176} {arXiv:1401.2176 [hep-lat]} \BibitemShut
  {NoStop}%
\bibitem [{\citenamefont {Brower}\ \emph {et~al.}(2014)\citenamefont {Brower},
  \citenamefont {Hasenfratz}, \citenamefont {Rebbi}, \citenamefont {Weinberg},\
  and\ \citenamefont {Witzel}}]{Brower:2014dfa}%
  \BibitemOpen
  \bibfield  {author} {\bibinfo {author} {\bibfnamefont {R.}~\bibnamefont
  {Brower}}, \bibinfo {author} {\bibfnamefont {A.}~\bibnamefont {Hasenfratz}},
  \bibinfo {author} {\bibfnamefont {C.}~\bibnamefont {Rebbi}}, \bibinfo
  {author} {\bibfnamefont {E.}~\bibnamefont {Weinberg}}, \ and\ \bibinfo
  {author} {\bibfnamefont {O.}~\bibnamefont {Witzel}},\ }\href@noop {} {\
  (\bibinfo {year} {2014})},\ \Eprint {http://arxiv.org/abs/1410.4091}
  {arXiv:1410.4091 [hep-lat]} \BibitemShut {NoStop}%
\bibitem [{\citenamefont {Cheng}\ \emph
  {et~al.}(2013{\natexlab{a}})\citenamefont {Cheng}, \citenamefont
  {Hasenfratz}, \citenamefont {Petropoulos},\ and\ \citenamefont
  {Schaich}}]{Cheng:2013eu}%
  \BibitemOpen
  \bibfield  {author} {\bibinfo {author} {\bibfnamefont {A.}~\bibnamefont
  {Cheng}}, \bibinfo {author} {\bibfnamefont {A.}~\bibnamefont {Hasenfratz}},
  \bibinfo {author} {\bibfnamefont {G.}~\bibnamefont {Petropoulos}}, \ and\
  \bibinfo {author} {\bibfnamefont {D.}~\bibnamefont {Schaich}},\ }\href
  {\doibase 10.1007/JHEP07(2013)061} {\bibfield  {journal} {\bibinfo  {journal}
  {JHEP}\ }\textbf {\bibinfo {volume} {1307}},\ \bibinfo {pages} {061}
  (\bibinfo {year} {2013}{\natexlab{a}})},\ \Eprint
  {http://arxiv.org/abs/1301.1355} {arXiv:1301.1355 [hep-lat]} \BibitemShut
  {NoStop}%
\bibitem [{\citenamefont {Cheng}\ \emph {et~al.}(2012)\citenamefont {Cheng},
  \citenamefont {Hasenfratz},\ and\ \citenamefont {Schaich}}]{Cheng:2011ic}%
  \BibitemOpen
  \bibfield  {author} {\bibinfo {author} {\bibfnamefont {A.}~\bibnamefont
  {Cheng}}, \bibinfo {author} {\bibfnamefont {A.}~\bibnamefont {Hasenfratz}}, \
  and\ \bibinfo {author} {\bibfnamefont {D.}~\bibnamefont {Schaich}},\ }\href
  {\doibase 10.1103/PhysRevD.85.094509} {\bibfield  {journal} {\bibinfo
  {journal} {Phys.Rev.}\ }\textbf {\bibinfo {volume} {D85}},\ \bibinfo {pages}
  {094509} (\bibinfo {year} {2012})},\ \Eprint {http://arxiv.org/abs/1111.2317}
  {arXiv:1111.2317 [hep-lat]} \BibitemShut {NoStop}%
\bibitem [{\citenamefont {Schaich}(2013)}]{Schaich:2013eba}%
  \BibitemOpen
  \bibfield  {author} {\bibinfo {author} {\bibfnamefont {D.}~\bibnamefont
  {Schaich}} (\bibinfo {collaboration} {USBSM}),\ }\href@noop {} {\bibfield
  {journal} {\bibinfo  {journal} {PoS}\ } (\bibinfo {year} {2013})},\ \Eprint
  {http://arxiv.org/abs/1310.7006} {arXiv:1310.7006 [hep-lat]} \BibitemShut
  {NoStop}%
\bibitem [{\citenamefont {Hasenfratz}\ \emph {et~al.}(2014)\citenamefont
  {Hasenfratz}, \citenamefont {Schaich},\ and\ \citenamefont
  {Veernala}}]{Hasenfratz:2014rna}%
  \BibitemOpen
  \bibfield  {author} {\bibinfo {author} {\bibfnamefont {A.}~\bibnamefont
  {Hasenfratz}}, \bibinfo {author} {\bibfnamefont {D.}~\bibnamefont {Schaich}},
  \ and\ \bibinfo {author} {\bibfnamefont {A.}~\bibnamefont {Veernala}},\
  }\href@noop {} {\  (\bibinfo {year} {2014})},\ \Eprint
  {http://arxiv.org/abs/1410.5886} {arXiv:1410.5886 [hep-lat]} \BibitemShut
  {NoStop}%
\bibitem [{\citenamefont {Hasenfratz}\ \emph {et~al.}(2007)\citenamefont
  {Hasenfratz}, \citenamefont {Hoffmann},\ and\ \citenamefont
  {Schaefer}}]{Hasenfratz:2007rf}%
  \BibitemOpen
  \bibfield  {author} {\bibinfo {author} {\bibfnamefont {A.}~\bibnamefont
  {Hasenfratz}}, \bibinfo {author} {\bibfnamefont {R.}~\bibnamefont
  {Hoffmann}}, \ and\ \bibinfo {author} {\bibfnamefont {S.}~\bibnamefont
  {Schaefer}},\ }\href {\doibase 10.1088/1126-6708/2007/05/029} {\bibfield
  {journal} {\bibinfo  {journal} {JHEP}\ }\textbf {\bibinfo {volume} {0705}},\
  \bibinfo {pages} {029} (\bibinfo {year} {2007})},\ \Eprint
  {http://arxiv.org/abs/hep-lat/0702028} {arXiv:hep-lat/0702028 [hep-lat]}
  \BibitemShut {NoStop}%
\bibitem [{\citenamefont {Cheng}\ \emph
  {et~al.}(2013{\natexlab{b}})\citenamefont {Cheng}, \citenamefont
  {Hasenfratz}, \citenamefont {Petropoulos},\ and\ \citenamefont
  {Schaich}}]{Cheng:2013bca}%
  \BibitemOpen
  \bibfield  {author} {\bibinfo {author} {\bibfnamefont {A.}~\bibnamefont
  {Cheng}}, \bibinfo {author} {\bibfnamefont {A.}~\bibnamefont {Hasenfratz}},
  \bibinfo {author} {\bibfnamefont {G.}~\bibnamefont {Petropoulos}}, \ and\
  \bibinfo {author} {\bibfnamefont {D.}~\bibnamefont {Schaich}},\ }\href@noop
  {} {\bibfield  {journal} {\bibinfo  {journal} {PoS}\ }\textbf {\bibinfo
  {volume} {LATTICE2013}},\ \bibinfo {pages} {088} (\bibinfo {year}
  {2013}{\natexlab{b}})},\ \Eprint {http://arxiv.org/abs/1311.1287}
  {arXiv:1311.1287 [hep-lat]} \BibitemShut {NoStop}%
\bibitem [{\citenamefont {Duane}\ \emph {et~al.}(1987)\citenamefont {Duane},
  \citenamefont {Kennedy}, \citenamefont {Pendleton},\ and\ \citenamefont
  {Roweth}}]{Duane:1987de}%
  \BibitemOpen
  \bibfield  {author} {\bibinfo {author} {\bibfnamefont {S.}~\bibnamefont
  {Duane}}, \bibinfo {author} {\bibfnamefont {A.}~\bibnamefont {Kennedy}},
  \bibinfo {author} {\bibfnamefont {B.}~\bibnamefont {Pendleton}}, \ and\
  \bibinfo {author} {\bibfnamefont {D.}~\bibnamefont {Roweth}},\ }\href
  {\doibase 10.1016/0370-2693(87)91197-X} {\bibfield  {journal} {\bibinfo
  {journal} {Phys.Lett.}\ }\textbf {\bibinfo {volume} {B195}},\ \bibinfo
  {pages} {216} (\bibinfo {year} {1987})}\BibitemShut {NoStop}%
\bibitem [{\citenamefont {Osborn}\ \emph {et~al.}()\citenamefont {Osborn} \emph
  {et~al.}}]{FUEL}%
  \BibitemOpen
  \bibfield  {author} {\bibinfo {author} {\bibfnamefont {J.}~\bibnamefont
  {Osborn}} \emph {et~al.},\ }\href {http://usqcd-software.github.io/FUEL.html}
  {\enquote {\bibinfo {title} {{Framework for unified evolution of lattices
  (FUEL)}},}\ }\BibitemShut {NoStop}%
\bibitem [{\citenamefont {Hasenbusch}\ and\ \citenamefont
  {Necco}(2004)}]{Hasenbusch:2004yq}%
  \BibitemOpen
  \bibfield  {author} {\bibinfo {author} {\bibfnamefont {M.}~\bibnamefont
  {Hasenbusch}}\ and\ \bibinfo {author} {\bibfnamefont {S.}~\bibnamefont
  {Necco}},\ }\href {\doibase 10.1088/1126-6708/2004/08/005} {\bibfield
  {journal} {\bibinfo  {journal} {JHEP}\ }\textbf {\bibinfo {volume} {0408}},\
  \bibinfo {pages} {005} (\bibinfo {year} {2004})},\ \Eprint
  {http://arxiv.org/abs/hep-lat/0405012} {arXiv:hep-lat/0405012 [hep-lat]}
  \BibitemShut {NoStop}%
\bibitem [{\citenamefont {Bilson-Thompson}\ \emph {et~al.}(2003)\citenamefont
  {Bilson-Thompson}, \citenamefont {Leinweber},\ and\ \citenamefont
  {Williams}}]{BilsonThompson:2002jk}%
  \BibitemOpen
  \bibfield  {author} {\bibinfo {author} {\bibfnamefont {S.~O.}\ \bibnamefont
  {Bilson-Thompson}}, \bibinfo {author} {\bibfnamefont {D.~B.}\ \bibnamefont
  {Leinweber}}, \ and\ \bibinfo {author} {\bibfnamefont {A.~G.}\ \bibnamefont
  {Williams}},\ }\href {\doibase 10.1016/S0003-4916(03)00009-5} {\bibfield
  {journal} {\bibinfo  {journal} {Annals Phys.}\ }\textbf {\bibinfo {volume}
  {304}},\ \bibinfo {pages} {1} (\bibinfo {year} {2003})},\ \Eprint
  {http://arxiv.org/abs/hep-lat/0203008} {arXiv:hep-lat/0203008 [hep-lat]}
  \BibitemShut {NoStop}%
\bibitem [{\citenamefont {Narayanan}\ and\ \citenamefont
  {Neuberger}(2006)}]{Narayanan:2006rf}%
  \BibitemOpen
  \bibfield  {author} {\bibinfo {author} {\bibfnamefont {R.}~\bibnamefont
  {Narayanan}}\ and\ \bibinfo {author} {\bibfnamefont {H.}~\bibnamefont
  {Neuberger}},\ }\href {\doibase 10.1088/1126-6708/2006/03/064} {\bibfield
  {journal} {\bibinfo  {journal} {JHEP}\ }\textbf {\bibinfo {volume} {0603}},\
  \bibinfo {pages} {064} (\bibinfo {year} {2006})},\ \Eprint
  {http://arxiv.org/abs/hep-th/0601210} {arXiv:hep-th/0601210 [hep-th]}
  \BibitemShut {NoStop}%
\bibitem [{\citenamefont {L{\"u}scher}(2010)}]{Luscher:2009eq}%
  \BibitemOpen
  \bibfield  {author} {\bibinfo {author} {\bibfnamefont {M.}~\bibnamefont
  {L{\"u}scher}},\ }\href {\doibase 10.1007/s00220-009-0953-7} {\bibfield
  {journal} {\bibinfo  {journal} {Commun.Math.Phys.}\ }\textbf {\bibinfo
  {volume} {293}},\ \bibinfo {pages} {899} (\bibinfo {year} {2010})},\ \Eprint
  {http://arxiv.org/abs/0907.5491} {arXiv:0907.5491 [hep-lat]} \BibitemShut
  {NoStop}%
\bibitem [{\citenamefont {L{\"{u}}scher}(2010)}]{Luscher:2010iy}%
  \BibitemOpen
  \bibfield  {author} {\bibinfo {author} {\bibfnamefont {M.}~\bibnamefont
  {L{\"{u}}scher}},\ }\href {\doibase 10.1007/JHEP08(2010)071} {\bibfield
  {journal} {\bibinfo  {journal} {JHEP}\ }\textbf {\bibinfo {volume} {1008}},\
  \bibinfo {pages} {071} (\bibinfo {year} {2010})},\ \Eprint
  {http://arxiv.org/abs/1006.4518} {arXiv:1006.4518 [hep-lat]} \BibitemShut
  {NoStop}%
\bibitem [{\citenamefont {Hasenfratz}(2014{\natexlab{a}})}]{AnnaTBP}%
  \BibitemOpen
  \bibfield  {author} {\bibinfo {author} {\bibfnamefont {A.}~\bibnamefont
  {Hasenfratz}},\ }\href@noop {} {\enquote {\bibinfo {title} {{Non-perturbative
  reduction of cut-off effects of the $t_0$, $w_0$ gradient flow scales}},}\ }
  (\bibinfo {year} {2014}{\natexlab{a}}),\ \bibinfo {note} {{In
  preparation}}\BibitemShut {NoStop}%
\bibitem [{\citenamefont {Hasenfratz}(2014{\natexlab{b}})}]{AnnaLat14Talk}%
  \BibitemOpen
  \bibfield  {author} {\bibinfo {author} {\bibfnamefont {A.}~\bibnamefont
  {Hasenfratz}},\ }\href
  {https://indico.bnl.gov/materialDisplay.py?contribId=100&sessionId=3&materialId=slides&confId=736}
  {\enquote {\bibinfo {title} {{Improved gradient flow for step scaling
  function and scale setting}},}\ } (\bibinfo {year} {2014}{\natexlab{b}}),\
  \bibinfo {note} {{Talk presented at Lattice 2014}}\BibitemShut {NoStop}%
\bibitem [{\citenamefont {Foley}\ \emph {et~al.}(2005)\citenamefont {Foley},
  \citenamefont {Jimmy~Juge}, \citenamefont {O'Cais}, \citenamefont {Peardon},
  \citenamefont {Ryan} \emph {et~al.}}]{Foley:2005ac}%
  \BibitemOpen
  \bibfield  {author} {\bibinfo {author} {\bibfnamefont {J.}~\bibnamefont
  {Foley}}, \bibinfo {author} {\bibfnamefont {K.}~\bibnamefont {Jimmy~Juge}},
  \bibinfo {author} {\bibfnamefont {A.}~\bibnamefont {O'Cais}}, \bibinfo
  {author} {\bibfnamefont {M.}~\bibnamefont {Peardon}}, \bibinfo {author}
  {\bibfnamefont {S.~M.}\ \bibnamefont {Ryan}},  \emph {et~al.},\ }\href
  {\doibase 10.1016/j.cpc.2005.06.008} {\bibfield  {journal} {\bibinfo
  {journal} {Comput.Phys.Commun.}\ }\textbf {\bibinfo {volume} {172}},\
  \bibinfo {pages} {145} (\bibinfo {year} {2005})},\ \Eprint
  {http://arxiv.org/abs/hep-lat/0505023} {arXiv:hep-lat/0505023 [hep-lat]}
  \BibitemShut {NoStop}%
\end{thebibliography}%
\bibliographystyle{apsrev4-1}
}


\end{document}